\documentclass[useAMS,usenatbib]{mn2e}

\def\Msun{\hbox{M$_\odot$}}
\def\Zsun{\hbox{Z$_\odot$}}
\def\Ne{\hbox{$n_{\rm {e}}$}}

\def\kms{\hbox{km$\,$s$^{-1}$}}
\def\cmt{\hbox{cm$^{-3}$}}
\def\Apix{\hbox{\AA$\,$pix$^{-1}$}}
\def\one{\,{\sc i}}             
\def\two{\,{\sc ii}}
\def\three{\,{\sc iii}}

\newcommand\fsec{\hbox{$.\!\!^{\rm s}$}}

\usepackage{amsmath}
\usepackage{amssymb}
\usepackage{mathrsfs}
\usepackage{color}
\usepackage{upgreek}
\usepackage[pdftex]{graphicx}

\defcitealias{westm07a}{Paper I}
\defcitealias{westm07b}{Paper II}
\title[The galactic outflow in NGC 1569]{Studying the galactic outflow in NGC 1569}
\author[M.S. Westmoquette et al.] {M.S. Westmoquette$^1$\thanks{E-mail:msw@star.ucl.ac.uk}, L. J. Smith$^{1,2}$ and J. S. Gallagher III$^3$\\
$^1$Department of Physics and Astronomy, University College London, Gower Street, London, WC1E 6BT\\
$^2$Space Telescope Science Institute and European Space Agency, 3700 San Martin Drive, Baltimore, MD 21218, USA\\
$^3$Department of Astronomy, University of Wisconsin-Madison, 5534 Sterling, 475 North Charter St., Madison WI 53706, USA\\
}
\date{Accepted. Received; in original form}
\pagerange{\pageref{firstpage}--\pageref{lastpage}}
\pubyear{2007}
\begin{document}
\maketitle
\label{firstpage}
\begin{abstract}
We present deep WIYN H$\alpha$ imaging of the dwarf irregular starburst galaxy NGC 1569, together with WIYN SparsePak spatially-resolved optical spectroscopy of the galactic outflow. This leads on from our previous detailed analyses of the state of the ISM in the central regions of this galaxy. Our deep imaging reveals previously undetected ionized filaments in the outer halo. Through combining these results with our spectroscopy we have been able to re-define the spatial extent of the previously catalogued superbubbles, and derive estimates for their expansion velocities, which we find to be in the range 50--100~\kms. The implied dynamical ages of $\lesssim$\,25~Myr are consistent with the recent star- and cluster-formation histories of the galaxy. Detailed decomposition of the multi-component H$\alpha$ line has shown that within a distinct region $\sim$$700\times 500$~pc in size, roughly centred on the bright super star cluster A, the profile is composed of a bright, narrow (FWHM $\lesssim$ 70~\kms) feature with an underlying, broad component (FWHM $\sim$ 150~\kms). Applying the conclusions found in our previous work regarding the mechanism through which the broad component is produced, we associate the faint, broad emission with the interaction of the hot, fast-flowing winds from the young star clusters with cool clumps of ISM material. This interaction generates turbulent mixing layers on the surface of the clouds and the evaporation and/or ablation of material into the outflow. Under this interpretation, the extent of the broad component region may indicate an important transition point in the outflow, where ordered expansion begins to dominate over turbulent motion. In this context, we present a multi-wavelength discussion of the evolutionary state of the outflow.
\end{abstract}

\begin{keywords} galaxies: individual (NGC 1569) -- galaxies: starburst -- galaxies: ISM -- ISM: kinematics and dynamics -- ISM: jets and outflows.
\end{keywords}

\section{Introduction}\label{intro}

Outflows powered by the collective injection of kinetic energy and momentum from massive stars and supernovae (SNe) can drastically affect the structure and subsequent evolution of galaxies. Thus, a good understanding of the feedback mechanisms between massive stars, star clusters and the ISM is fundamental. In particular, dwarf galaxies are thought to be strongly affected by the effects of feedback since their smaller gravitational potentials mean that supernova-heated gas can escape more easily \citep{larson74}. Although the ejection of the ISM potentially could have significant consequences on the star-formation rate within these low-mass systems \citep{dekel86}, more recent work suggests that ejection of hot gas through bubble blow-out may not be as efficient as first thought \citep{deyoung94, martin98}. It is therefore important to study such systems to understand how gas is removed and what effects this has in the evolution of the galaxy.

NGC 1569 (UGC 3056, Arp 210, VII Zw 16, IRAS 4260+6444) is a nearby \citep[2.2~Mpc;][]{israel88}, low metalicity \citep*[0.25~\Zsun;][]{devost97, kobulnicky97} dwarf irregular galaxy that has recently undergone a period of starburst activity. The most recent burst is thought to have peaked between $\sim$10--100~Myr ago with an average star-formation rate of $\sim$0.5~\Msun~yr$^{-1}$ \citep{greggio98}. At some point near the end of this event, the two well-known super-star clusters (SSCs) A and B were formed (\citealt{arp85}; \citealt*{oconnell94}; \citealt{demarchi97}), and together with the slightly older cluster 30 \citep{hunter00, origlia01}, appear to dominate the energetics of the central regions.

H\one\ observations of NGC 1569 \citep{stil98, stil02, muhle05} clearly show morphological and kinematical signatures caused by the starburst. Firstly, the H\one\ kinematics are ``strongly disrupted'' within the central 900~pc, with little or no evidence for the disc rotation seen at lager radii. Furthermore, large seemingly tidal structures are seen, including a so-called bridge connecting the galaxy to a low-mass H\one\ cloud \citep[the companion;][]{stil98}, a large H\one\ arm extending to the west of the disc, and a very faint filamentary H\one\ stream wrapping around the south of the disc \citep{muhle05}. A `hot spot' (a region of velocity crowding) on the western edge of the disk was found by \citet{muhle05}, who interpreted it as the impact location of infalling gas from the companion. This provides a compelling explanation of how the starburst event was triggered. 

H$\alpha$ images of this galaxy show an equally chaotic, complex structure to the warm ionized component, exhibiting many filaments, bubbles and loops. Many of these were identified by \citet*{hunter93} from deep H$\alpha$ imaging. Later kinematical studies found that these filaments form part of a cellular outflow structure, comprising large-scale expanding superbubbles on the northern and southern sides of the disc (\citealt*{tomita94}; \citealt{heckman95, martin98}). By analysing spectra from two long-slits placed parallel to the major and minor axes, \citet{heckman95} found evidence of shocks in the outer regions of the ionized halo. The ratios of [O\one]/H$\alpha$, [S\two]/H$\alpha$ and [N\two]/H$\alpha$ \citep[used often to diagnose and trace the conditions within ionized gas;][]{veilleux87, dopita00} were all found to be high in the outer filaments, indicating either an increase in the importance of shocks, or that photoionization becomes less important with distance as the ionizing radiation from the central starburst becomes diluted.

The existence of shocked gas is supported by high-resolution X-ray observations of NGC 1569. \citet*{martin02} examined the X-ray properties of the outflow with \textit{Chandra}, and found significant soft (0.3--0.7~keV), diffuse emission coincident with the H$\alpha$ morphology. Although they found the X-ray colour variations to be inconsistent with a free-streaming wind, they concluded that the X-ray emission probably originates in the halo shock generated by the outflowing gas, possibly from the mixing layers between the shock and the bubble interior.

In order to properly characterise the structure of the outflow, it is essential to study its kinematics. From optical long-slit spectroscopy, \citet{martin98} found the outflow to be composed of numerous superbubbles. In general, she finds the redshifted component of the split-line profile to be stronger in the south and weaker in the north, suggesting an inclined outflow aspect. This is consistent with more recent X-ray absorption measurements \citep{martin02} and H\one\ observations \citep[from which an inclination angle of $\sim$60$^{\circ}$ was derived][]{stil02}. Although it is unclear whether the two SSCs, A and B, alone are sufficiently powerful to provide enough mechanical energy and ionizing radiation to drive the whole outflow and power the galaxy's diffuse ionized medium \citep{martin97, martin02}, what is clear from the H$\alpha$ morphology is that energy is being injected throughout the central starburst zone of the disc from multiple sources.

A detailed look at the spectral line profiles, however, shows that near SSC A, ``distinctly non-Gaussian'' H$\alpha$ emission can be found, exhibiting weak but very broad wings \citep{heckman95}. Broad emission line wings have been detected in other nearby starburst galaxies (e.g.~\citealt{izotov96, homeier99, marlowe95, mendez97}; \citealt*{sidoli06}; \citealt{westm07c}). Due to mismatches in spectral and spatial resolution and in the specific environments observed, the nature of the energy source for these broad lines has been contested, and has resulted in the proposal of number of possible explanations. However, a detailed IFU (integral field unit) study of the ionized ISM conditions in four regions within the 200~pc region surrounding SSC A by \citet[][hereafter \citetalias{westm07a}]{westm07a} and \citet[][hereafter \citetalias{westm07b}]{westm07b} have shed a considerable amount of light on this problem.

By accurately decomposing the emission line profiles across each of the IFU fields, we found the line shape to be, in general, composed of a bright narrow feature (intrinsic FWHM $\sim$ 50~\kms) superimposed on a fainter broad component (FWHM $\sim$ 200--400~\kms). By mapping out their individual properties, we identified a number of correlations between the line components that allowed us to investigate in detail the state of the ionized ISM. We concluded that the broad underlying component is most likely produced in a turbulent mixing layer \citep{slavin93, binette99} on the surface of cool gas knots, set up by the impact of the fast-flowing cluster winds \citep{pittard05}. Our analysis revealed a very complex environment with many overlapping and superimposed components, but surprisingly no evidence for organised bulk motions. We concluded that the four regions sampled are all located well within the wind energy injection zone \citep{shopbell98} at the very roots of the outflow, and that the collimation processes required to transform the turbulent motions into an organised net outflow forming the large-scale superbubbles must occur between 100--200~pc from the central star clusters.

With this in mind, we have obtained new deep H$\alpha$ imaging of NGC 1569, together with spatially-resolved spectra of the outer halo, to investigate in detail the morphology and kinematics (including the line profile shapes) of the warm ionized component of the outflow at these large radii. In Section~\ref{sect:data} we present the observations and in Section~\ref{sect:maps} we map out the properties of the line components and discuss the results of the spectroscopy. Since a number of SparsePak fibres are coincident with or lie adjacent to the Gemini GMOS/IFU fields presented in Papers I and II, we compare the results obtained with the two instruments in Section~\ref{sect:comparison}. In Section~\ref{sect:disc} we discuss the state of the outflow, including the conditions in the inner and outer halo, and we summarise our findings and conclusions in Section~\ref{sect:concs}.

\section{Observations and Data Reduction} \label{sect:data}

\begin{figure*}
\centering
\includegraphics[width=\textwidth]{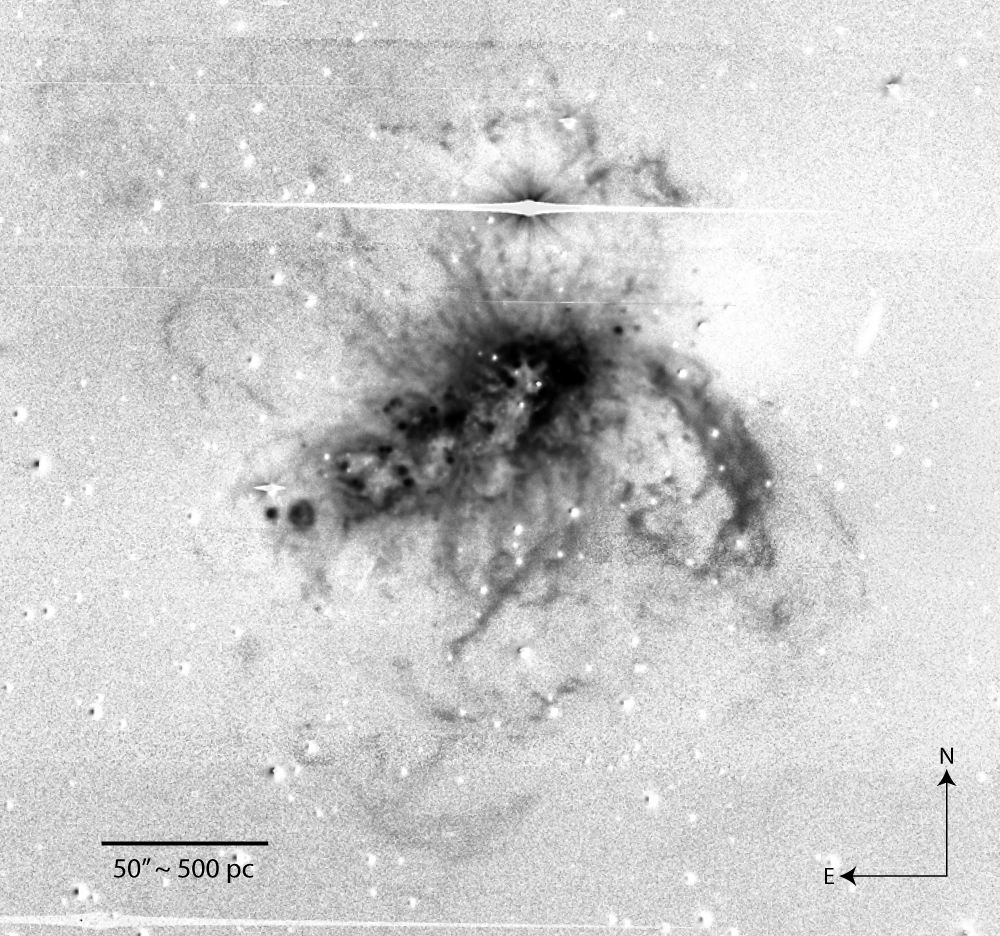}
\caption{Continuum subtracted H$\alpha$ image of NGC 1569 taken with WIYN/MiniMo. The intensity stretch has a cube-root scaling and is inverted to enhance low surface-brightness features.}
\label{fig:wiynha}
\end{figure*}

\begin{figure*}
\centering
\includegraphics[width=0.8\textwidth]{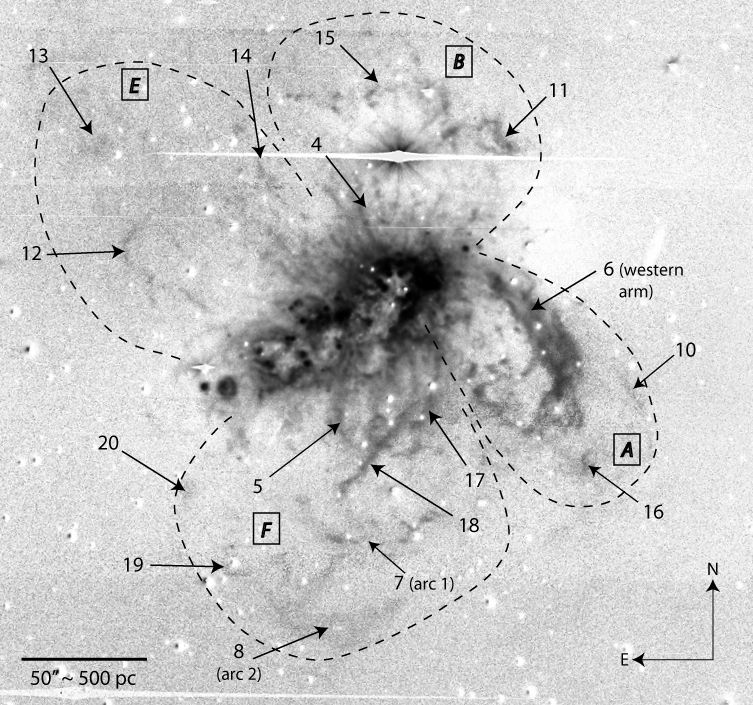}
\caption{H$\alpha$ morphology of the NGC 1569 halo. Numbered filaments correspond to the nomenclature of \citet{hunter93}; newly identified filaments are numbered from 14 onwards. The arc 1/arc 2 nomenclature is from \citet{waller91}. The dashed regions labelled with letters correspond to large-scale superbubbles and follow the nomenclature of \citet{martin98}. The white cross marks the position of SSC A.}
\label{fig:shells}
\end{figure*}

\subsection{Imaging} \label{sect:imaging}
We obtained WIYN\footnote{The WIYN Observatory is a joint facility of the University of Wisconsin-Madison, Indiana University, Yale University, and the National Optical Astronomy Observatories.} 3.5-m Mini-Mosaic (MiniMo) imaging of NGC 1569 on 13th December 2006 through Harris $R$-band and W15 ($\equiv$ H$\alpha$) filters, in seeing conditions of $\sim$$1\farcs1$--$1\farcs2$. Mini-Mos consists of two $4096\times 2048$ CCDs separated by a small gap. Two offset exposures were taken for each filter in order to cover the gap and to aid cosmic ray and defect removal; exposure times were 300~s for each $R$-band image and 1200~s for each H$\alpha$ image. Dome flats and bias frames were also taken. Reduction was achieved using the NOAO \textsc{mscred} package within \textsc{iraf}\footnote{The Image Reduction and Analysis Facility ({\sc iraf}) is distributed by the National Optical Astronomy Observatories which is operated by the Association of Universities for Research in Astronomy, Inc.\ under cooperative agreement with the National Science Foundation.}. First, bias and flat fields were combined and subtracted/divided into the object frames, then an accurate world coordinate system was applied using the tasks \textsc{msczero} and \textsc{mscmatch}. Single images were constructed from the individual chips using \textsc{mscimage}, and large-scale gradients in the sky background were removed using \textsc{mscskysub}. Finally the separate mosaiced frames were combined using \textsc{mscstack}, after matching their overall intensities using \textsc{mscimage}. The final mosaiced $R$ and H$\alpha$ images cover a field-of-view of $9.5\times 10.7$~arcmins at a spatial sampling of $0\farcs14$~pixel$^{-1}$.

The $R$-band image was used as a proxy off-band exposure to subtract the continuum from the H$\alpha$ image. The final continuum-subtracted H$\alpha$ mosaic is shown in Fig.~\ref{fig:wiynha}. Although the depth of this image equals that of the deepest image previously published \citep{hunter93}, its quality (particularly around the bright saturated star in the north of the image) and dynamic range allow previously undetected features to be seen. For the first time, the morphological structure of the H$\alpha$-emitting gas can be easily traced from the outer halo right into the central regions, where multiple nested bubbles and shells are organised in a seemingly chaotic manner. Sub-structure can be seen down to the resolution limit of the observations, and emission is discernible out to distances of 2.5--3~arcmins (1500--1800~pc) from the disc. The halo morphology seen in Fig.~\ref{fig:wiynha} is clearly very complex.

\subsubsection{Shell morphology}
Kinematical studies of the NGC 1569 halo have identified a number of large-scale supershells through their double-peaked signatures in bright emission lines. Through these methods, \citet{martin98} found the southern flow to be composed of three main expanding shells (which she denoted F, G, and A). The largest line-splitting was found in shell A, outlined on the western side by the prominent `western arm' \citep{hodge74}, corresponding to expansion velocities of $v_{\rm exp} \sim 70$--100~\kms\, and indicating that this bubble might not be ``strongly decelerated as it pushes through galactic halo''. The northern halo was also found to be composed of three large-scale shells (E, D and B) where the expansion velocity of shell B was found to be $v_{\rm exp}\sim 80$~\kms. Shell C, also identified by \citet{tomita94} in their stepped long-slit observations, represents the central cavity in NGC 1569 centred on SSCs A and B, and measures $\sim$$20''$ in diameter with a $v_{\rm exp}$ of $\sim$40~\kms.

Deep H$\alpha$ imaging reveals a cornucopia of structure within the halo that was first catalogued by \citet{hunter93}. The depth and quality of our new imaging (Fig.~\ref{fig:wiynha}) shows many more features that can now be added to the list. In Fig.~\ref{fig:shells} we have labelled the prominent H$\alpha$ halo features following the nomenclature of \citet{hunter93}, where new detections are numbered from 14 onwards. To clarify matters, we suggest that: filament 4 refers to the whole set of linear streamers emanating from the north-west of the disc; filament 5 refers to the bubble-like curved filament, whereas filament 18 refers to the brighter linear streamer; and filament 6 refers to the whole of the western arm.

Based on the imaging and spectroscopy presented in this paper, we have re-examined the existence and extent of the large-scale superbubbles catalogued by \citet{martin98}. Previously undetected faint structure and spectroscopically identified expanding material (see Section~\ref{sect:maps}) have allowed us to refine the spatial extents of supershells A and B. However, the H$\alpha$ morphology does not support a distinction between shells G and F \citep{martin98}, so we propose that the whole south-eastern shell complex consisting of filaments 5, 7, 8, 17, 18, 19, and 20 should be referred to as superbubble F. This bubble corresponds to the south-eastern X-ray lobe \citep[][and Section~\ref{sect:outer}]{martin02}. In our interpretation, the two kinematically identified expanding structures that led \citet{martin98} to separately define shells G and F, represent just one level of a `hierarchy of structure', where multiple nested bubbles (e.g.~19 and 8 or 17 and 5) are continuously interacting to form larger structures. This interpretation can also be applied to the north-eastern halo, where a distinction between shells D and E is equally difficult to corroborate from the H$\alpha$ and X-ray morphologies (see Section~\ref{sect:outer}). Thus we suggest superbubble E should encompass the whole north-eastern halo, including both the faint H$\alpha$ wisps in this region (evident from both the H$\alpha$ imaging and spectroscopy presented below) and the extent of the north-eastern X-ray lobe \citep[][and Section~\ref{sect:outer}]{martin02}. The spatial extent of the large-scale superbubbles according to our interpretation are drawn on Fig.~\ref{fig:shells} with dashed lines and are labelled with the letters A, B, E and F.

The inclination of the disc \citep[$\sim$60$^{\circ}$;][]{stil02} is such that the northern side is tipped away from our sightline. This scenario is supported by the consistent reversal of line strengths between the north and south \citep{heckman95, martin98} and the more homogeneous H\one\ distribution in the north \citep{muhle05}. Therefore, unlike the southern side where we can, in effect, see up into the disc, the northern side is much more obscured. A large number of linear streamers (labelled collectively as filament 4) all point away from the central region containing SSCs A and B, and appear to form the tails of the swept-back, cometary-like shape of the gas knots seen just to the north of SSCs A and B \citepalias{westm07b}. These radial H$\alpha$ filaments could be compressed regions of shells or ablation trails from dense gas clouds \citep[e.g.][]{melioli05}.

\subsection{Spectroscopy} \label{sect:spec}
We observed four regions surrounding NGC 1569 with the SparsePak instrument \citep{bershady04} on the WIYN telescope during the period 13--16th December 2004. SparsePak is a ``formatted field unit'' similar in design to a traditional IFU, except that its 82 fibers are arranged in a sparsely packed grid, with a small, nearly-integral core \citep[see][their fig.~2 for a map of the fibre numbers]{bershady04}. SparsePak was designed to maximise throughput and spectral resultion at the expense of spatial coverage/resolution, and as such has a total light throughput of $\sim$90 per cent longwards of 500~nm. Each fibre has a diameter of 500~$\upmu$m, corresponding to $4\farcs69$ on the sky; the formatted field has approximate dimensions of $72\times 71.3$~arcsecs, including seven sky fibres located on the north and west side of the main array separated by $\sim$$25''$. The mapping order of fibers between telescope and spectrograph focal planes was purposefully designed in a fairly randomised fashion in order to distribute sky fibres evenly along the slit and minimise the effects of spectrograph vignetting on the summed sky spectrum. SparsePak is connected to the Hydra bench-mounted echelle spectrograph which uses a $2048\times 2048$ CCD detector. Using an order 8 grating with an angle of 63.25$^{\circ}$ (giving a spectral coverage of 6450--6865~\AA{} and dispersion of 0.20~\Apix), we were able to cover the nebular emission lines of H$\alpha$, [N\two]$\lambda\lambda 6548,6583$, and [S\two]$\lambda\lambda 6716,6731$. A number of bias frames, flat-fields and arc calibration exposures were also taken together with the science frames.

\begin{figure*}
\centering
\includegraphics[width=0.8\textwidth]{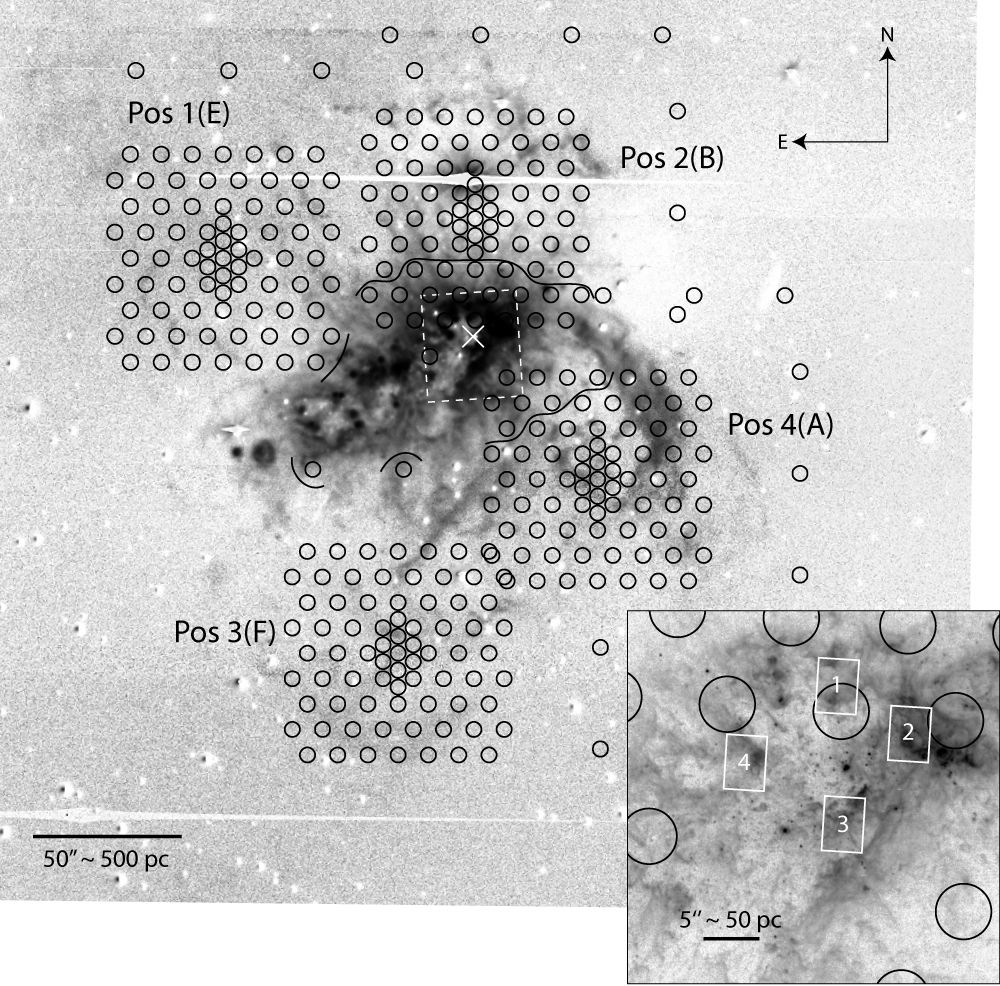}
\caption{WIYN H$\alpha$ image of NGC 1569 with the positions of the four SparsePak fields (including sky fibres) overlaid. The letters in brackets indicate the main supershell that they cover. The bold lines indicate the extent of the broad component region as described in Section~\ref{sect:inner}. The white cross marks the position of SSC A. \textit{Inset:} \textit{HST}/WFPC2 F656N image covering the white dashed area outlined on the main image, showing the position of the Gemini GMOS/IFU fields of view \citepalias[$5\times 3.5$~arcsecs;][]{westm07a,westm07b} on an \textit{HST}/WFPC2 F656N image, compared to the nearest SparsePak fibres.}
\label{fig:finder}
\end{figure*}

The positions of the SparsePak fibres for all four fields are shown in Fig.~\ref{fig:finder}, overlaid on the WIYN H$\alpha$ image. For clarity, we have named each position with a number, and a bracketed letter referring to the name of the supershell with which the position is associated. The arrangement of the fibres (including the offset sky fibres) can clearly be seen. A list of the coordinates for each position, and the total exposure times are given in Table~\ref{tbl:sp_obs}. The four pointings were chosen to sample the outer galactic wind flow, and coincide with the main expanding shells identified by \citet{heckman95} and \citet{martin98} and with regions of bright diffuse X-ray emission \citep{martin02}.

\begin{table*}
\centering
\caption {WIYN SparsePak observations}
\label{tbl:sp_obs}
\begin{tabular}{lccll}
\hline
Position & RA & Dec & Exposure Time & Sky fibres used \\
& \multicolumn{2}{c}{(J2000)} & (s) & for subtraction \\
\hline
1(E) & $4^{\rm h} 31^{\rm m} 01\fsec0$ & $64^{\circ} 51' 25\farcs0$ & $8\times 1800$ & 16,22 \\
2(B) & $4^{\rm h} 30^{\rm m} 48\fsec5$ & $64^{\circ} 51' 37\farcs0$ & $9\times 1800$ & 2,37,54,80 \\
3(F) & $4^{\rm h} 30^{\rm m} 52\fsec0$ & $64^{\circ} 49' 18\farcs0$ & $8\times 1800$ & 70,80 \\
4(A) & $4^{\rm h} 30^{\rm m} 42\fsec0$ & $64^{\circ} 50' 15\farcs0$ & $8\times 1800$ & 2,37,54,80 \\
\hline
\end{tabular}
\end{table*}

Basic reduction was achieved using the NOAO {\sc hydra} package within {\sc iraf}. The first step was to run {\sc apfind} on the flat-field exposure to automatically detect the individual spectra on the CCD frame. The output of this task is an aperture identification table which can be used for each science frame to extract out the individual spectra. The task {\sc dohydra} was then used to perform the bias subtraction, flat-fielding and wavelength calibration. The datafiles at this stage contained 82 reduced and wavelength calibrated spectra, each 1 pixel in width and ordered by the position they were recorded on the CCD (the `fibre order'). Because of the positioning of our fields on NGC 1569, many of the sky fibres fell on regions of strong nebular emission (see Fig.~\ref{fig:finder}) and were therefore unusable. By a combination of visual inspection of both the finding chart and the sky spectra themselves, we determined which fibres to use for the sky subtraction process (those that were contaminated were used as normal science fibres). A list of which fibres were used for sky subtraction in each position is given in Table~\ref{tbl:sp_obs}. After determination of the appropriate usable sky fibres, an average sky spectrum was created by averaging the spectra and sky-subtraction was achieved by re-running {\sc dohydra} with the sky-subtraction option switched on. Cosmic-rays were cleaned from the data using {\sc lacosmic} \citep{vandokkum01}, before final combination of the individual frames was achieved using {\sc imcombine}. The final datafiles now each contained 82 reduced, wavelength calibrated and sky-subtracted spectra.

In order to determine an accurate measurement of the instrumental contribution to the spectrum broadening, we selected high S/N spectral lines from a wavelength calibrated arc exposure that were close to the H$\alpha$ line in wavelength, and sufficiently isolated to avoid blends. After fitting these lines with Gaussians in all 82 apertures, we find the average instrumental width is $30.9\pm 0.4$~\kms.

\subsubsection{Line fitting} \label{sect:line_fitting}

\begin{figure*}
\centering
\begin{minipage}{7cm}
\includegraphics[width=7cm]{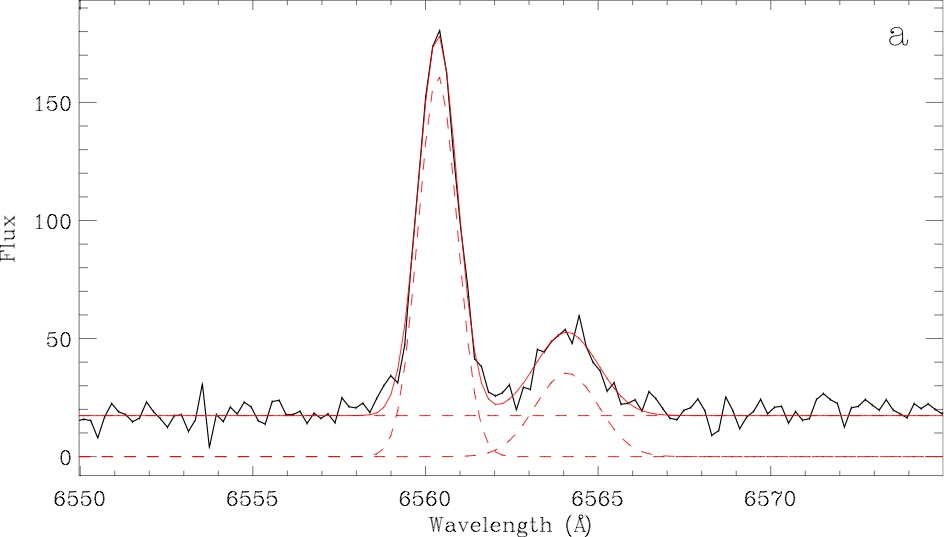}
\end{minipage}
\hspace{0.2cm}
\begin{minipage}{7cm}
\includegraphics[width=7cm]{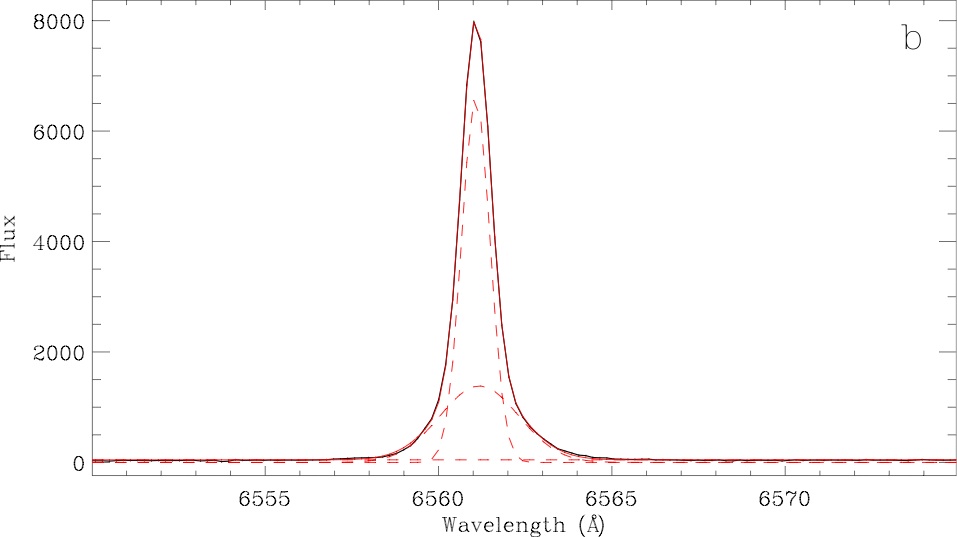}
\end{minipage}
\begin{minipage}{7cm}
\vspace{0.3cm}
\includegraphics[width=7cm]{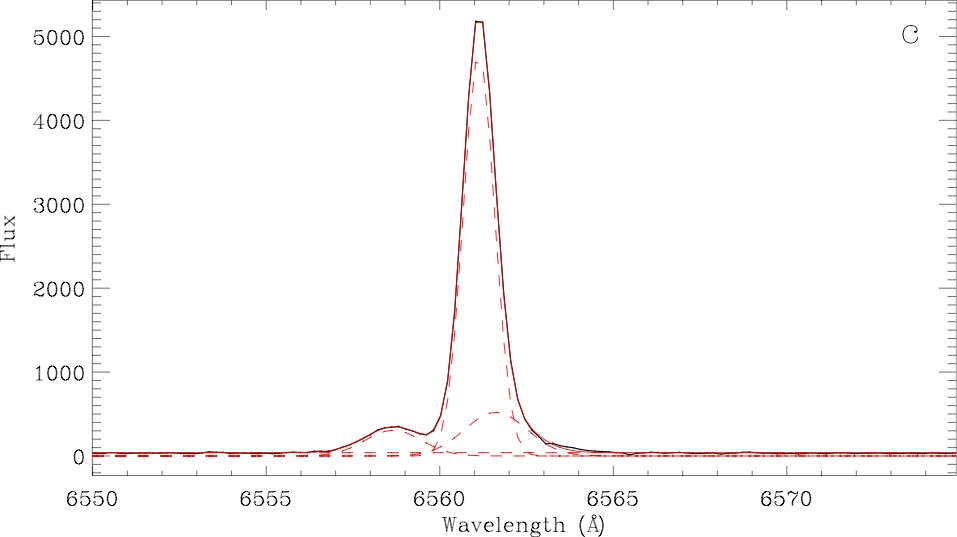}
\end{minipage}
\hspace{0.2cm}
\begin{minipage}{7cm}
\vspace{0.3cm}
\includegraphics[width=7cm]{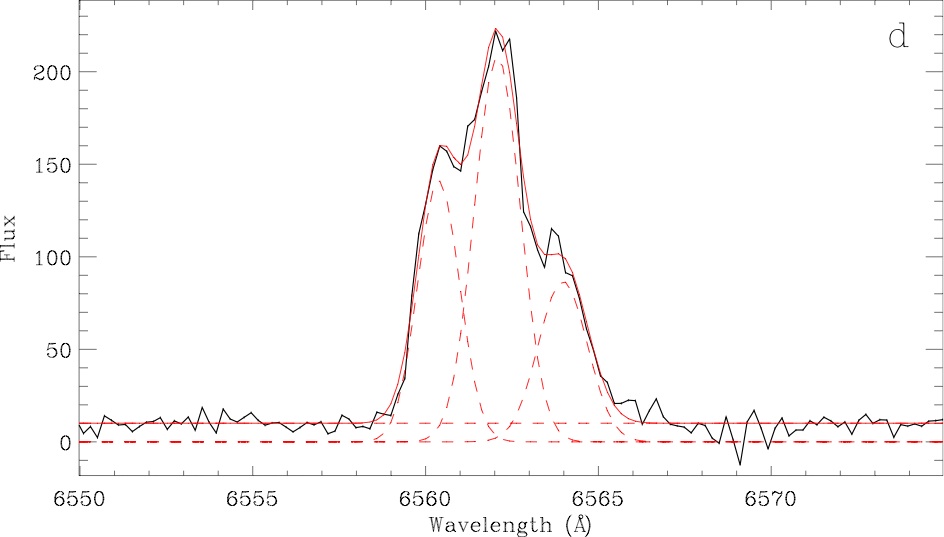}
\end{minipage}
\begin{minipage}{7cm}
\vspace{0.3cm}
\includegraphics[width=7cm]{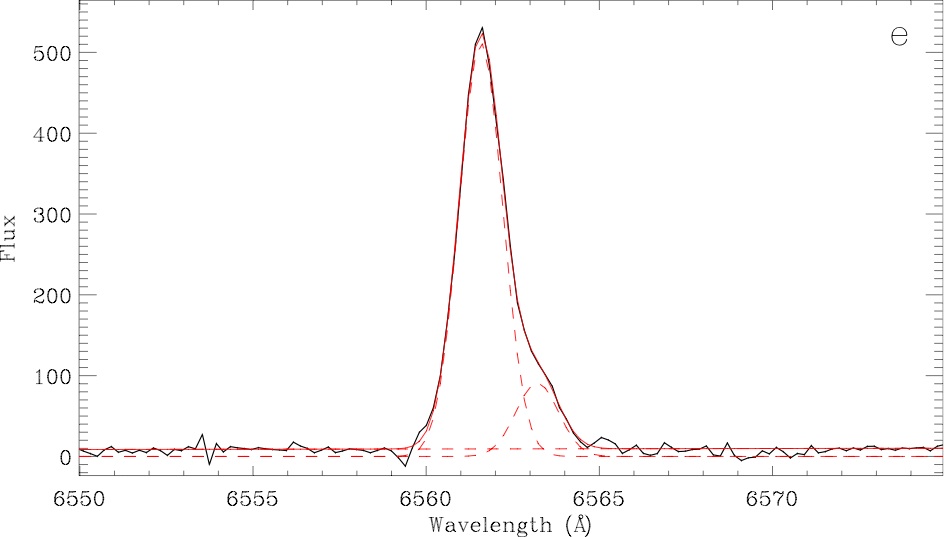}
\end{minipage}
\hspace{0.2cm}
\begin{minipage}{7cm}
\vspace{0.3cm}
\includegraphics[width=7cm]{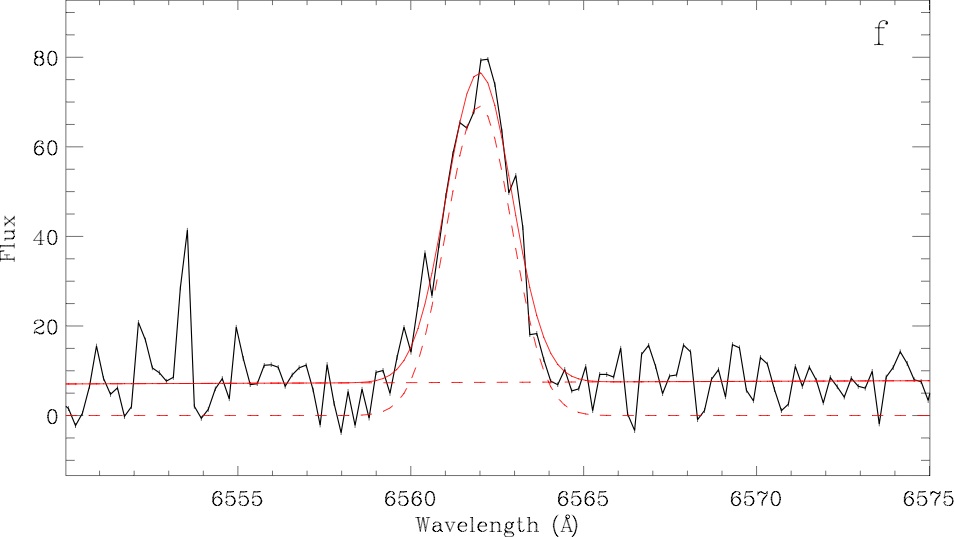}
\end{minipage}
\vspace{0.4cm}
\caption{Example H$\alpha$ line profiles chosen to represent the main types of profile shapes observed over the four fields (top four) and to aid explanation in the text (lower two). (a) split line profile; (b) narrow, bright component with underlying, broad component; (c) split narrow profiles with underlying broad component [from N-E of position 4(A)]; (d) triple component line [from centre of shell B, position 2(B)]; (e) a spaxel from near the south-east of position 2(B) ($30''$ from SSC A) that does not show a broad underlying component, only two narrow components; (f) a low S/N line broadened by multiple unresolved components (from the north-west of position 2(B) on the edge of shell B). Observed data is shown by a solid black line, individual Gaussian fits by dashed red lines (including the straight-line continuum level fit), and the summed model profile in solid red. Flux units are arbitrary but on the same scale.}
\label{fig:sp_eg_fits}
\end{figure*}

We fitted model Gaussian profiles to each of the emission lines detected in each spaxel of the four SparsePak fields using the $\chi^2$ fitting routine \textsc{pan} \citep[Peak ANalysis;][]{dimeo}. A detailed description of the program and the customisations we have made to it are given in \citetalias{westm07a}. In order to maintain a consistent approach to help in the minimisation of the fit and in the analysis, the first component of the fit initial guess (hereafter referred to as C1) was always assigned to the brightest of any multiple-components encountered. In many cases, a second Gaussian component (hereafter C2) was needed; sometimes it was required to fit a broad underlying profile, and in others it was required to fit the fainter component of a split line. The number of components fitted to each line was determined using a combination of visual inspection and the $\chi^{2}$ fit statistic output by \textsc{pan}. If, as was often found in the outer parts of the galaxy where the S/N of our spectra were lower, only one component was detected, but its velocity was consistent with the surrounding C2 velocities, we manually re-assigned the component to C2 in order to aid interpretation.

A range of example H$\alpha$ line profiles from various spaxels are shown in Fig.~\ref{fig:sp_eg_fits}, together with the individual Gaussian fits required to model the integrated line-shapes. The top four profiles were chosen to illustrate the main types of H$\alpha$ line shapes encountered over the four fields; the majority are similar to the top two examples, being composed of single- or double-Gaussians (most being simply double-peaked -- panel a -- but a good proportion exhibiting a broad underlying component -- panel b). In only five fibres do we find a clear triple profile in H$\alpha$, two of these being of the type shown in Fig.~\ref{fig:sp_eg_fits}c (three distinct narrow components), and three being of the type shown in panel d (broad underlying and faint secondary narrow component). The bottom two panels (e and f) show example profiles that are discussed in the following sections.

\section{Emission Line Maps} \label{sect:maps}

\begin{figure*}
\centering
\includegraphics[width=0.8\textwidth]{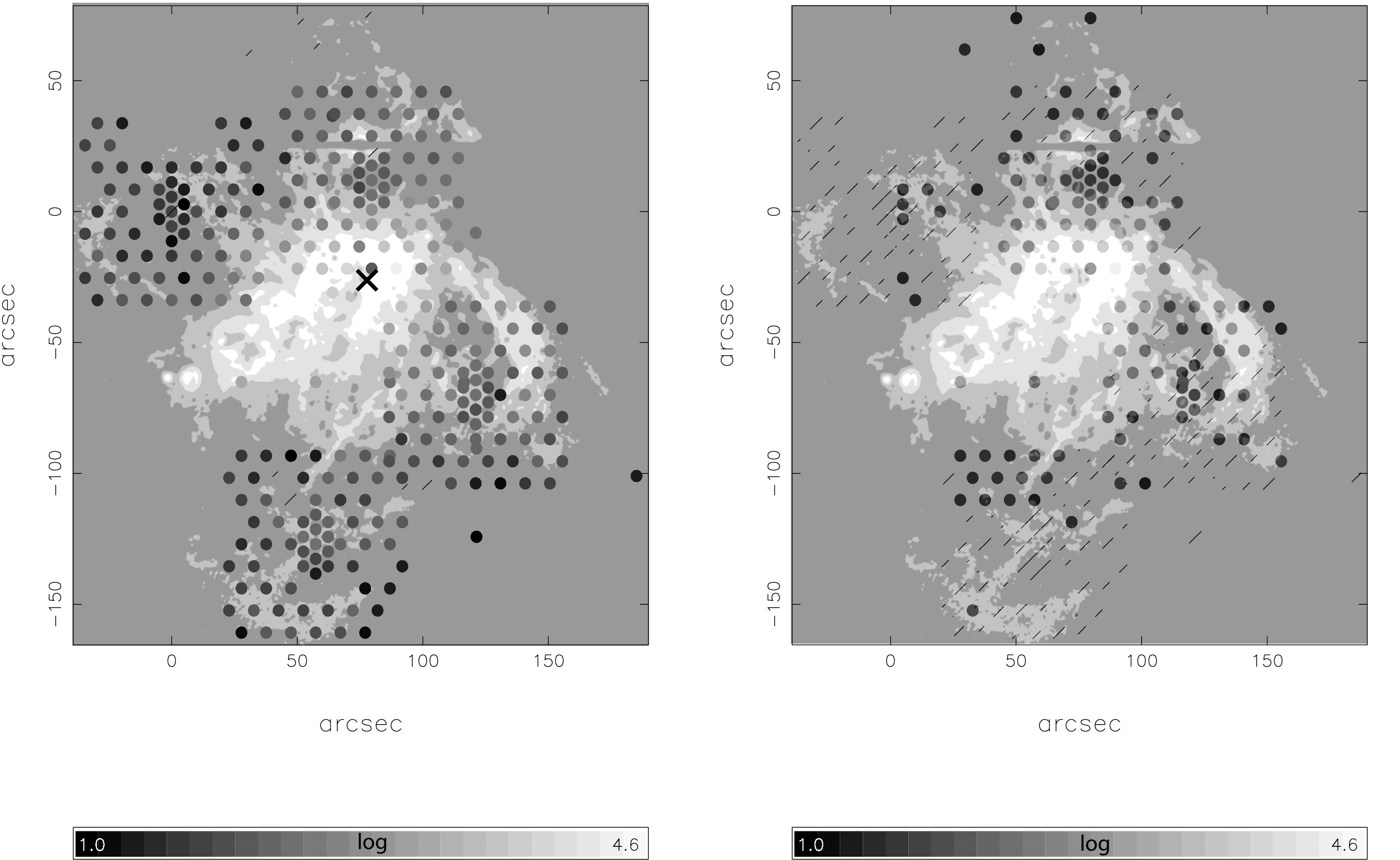}
\caption{H$\alpha$ flux maps for the four SparsePak fields for C1 and C2 (left and right panels respectively). A scale bar is given for each plot (log units, arbitrary scale). Hatched spaxels represent non-detections, and the origin is the field centre of position 1(E). North is up and east is left. A reproduction of the WIYN H$\alpha$ image is shown underneath the C1 map for means of comparison.}
\label{fig:sp_flux}
\end{figure*}

\begin{figure*}
\centering
\includegraphics[width=0.8\textwidth]{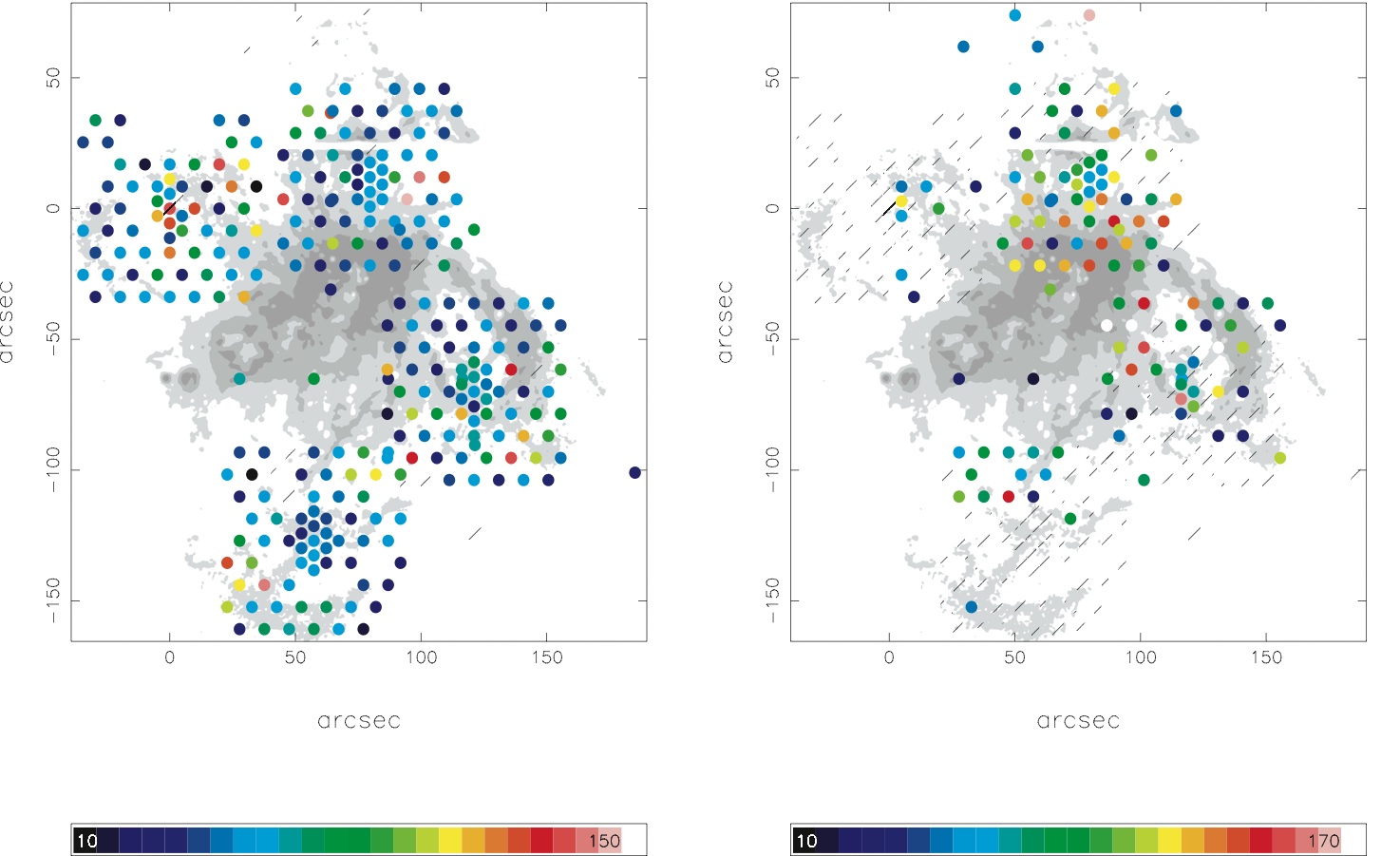}
\caption{FWHM maps of the H$\alpha$ line for the four SparsePak fields. \emph{Left:} C1 and \emph{right:} C2. A scale bar is given for each plot in units of \kms, corrected for instrumental broadening.}
\label{fig:sp_fwhm}
\end{figure*}

\begin{figure*}
\centering
\includegraphics[width=0.8\textwidth]{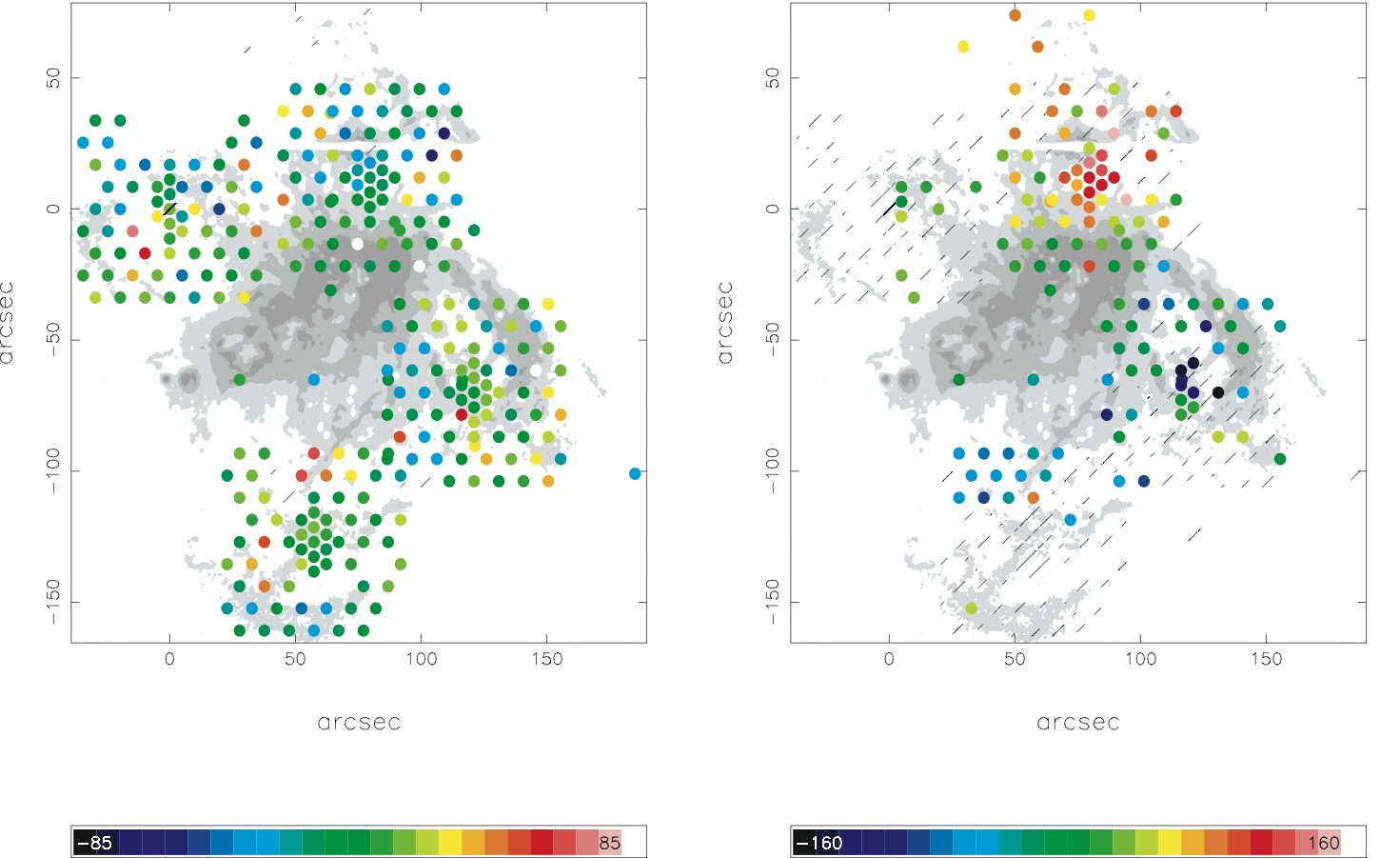}
\caption{Radial velocity maps of the H$\alpha$ line for the four SparsePak fields. \emph{Left:} C1 and \emph{right:} C2. A scale bar is given in units of \kms, relative to $v_{\rm sys}$, where the vertical lines denote a velocity of 0.}
\label{fig:sp_vel}
\end{figure*}

The spatial distribution of the properties of each Gaussian component identified in the SparsePak fields were mapped using the visualisation tool, {\sc daisy}, described in \citetalias{westm07a}. Figs~\ref{fig:sp_flux}, \ref{fig:sp_fwhm} and \ref{fig:sp_vel} show respectively the flux, FWHM and radial velocity maps of the H$\alpha$ line for the four SparsePak positions, overlaid on a faded reproduction of the WIYN H$\alpha$ image. For clarity, we do not include the third component since it is only detected in a few spaxels. In C1 (the brightest component) we detect H$\alpha$ emission out to the extent of our spatial coverage -- a maximum of 2.4 arcmins (1.5~kpc) away from SSC A.


Fig.~\ref{fig:sp_flux} shows the flux maps of C1 (left panel) and C2 (right panel) seen by SparsePak. It is clear that the flux variations of both components correspond well to the H$\alpha$ image shown in the background. The average flux ratio of C2/C1 over the four fields is $\sim$$0.7\pm 0.4$. This value gives an estimation of the importance of the different components; unfortunately the quality of these datasets is not high enough to reveal any trends in this value across the observed areas. Our derived C2/C1 ratio is considerably higher than that found from any of our GMOS IFU fields of the central regions \citepalias[0.1--0.3;][]{westm07b}. The reasons for this will be discussed in Section~\ref{sect:comparison}.

In Fig.~\ref{fig:sp_fwhm}, we plot the spatial variation of the FWHM of H$\alpha$. The line width of C1 (left-hand panel) remains fairly constant across the face of the galaxy, ranging over $\sim$20--70~\kms\ (corrected for the instrumental contribution). In a number of places broad C1 lines are measured, but because these are almost all located adjacent or near to spaxels where multiple components are detected, these broad fits are likely to be caused by the convolution of $\sim$two unresolved profiles. An example of a broad line from one of the outer-wind regions is shown in Fig.~\ref{fig:sp_eg_fits}f, illustrating how the low S/N has meant that only a single-Gaussian is required for a satisfactory fit. However, the width of this Gaussian suggests that the profile contains multiple kinematic components. The FWHM map for C2 is shown in the right-hand panel of Fig.~\ref{fig:sp_fwhm}. At large radii, where a second component is detected it typically has a width similar to that of C1, and forms part of a classic split-line profile (e.g.~Fig~\ref{fig:sp_eg_fits}a). The broadest lines (FWHM $\lesssim$ 160~\kms) are found within the central regions of the galaxy. Here C2 represents an underlying broad component of the type seen in Fig.~\ref{fig:sp_eg_fits}b, and very similar to what we found near the centre of the galaxy from our GMOS IFU study \citepalias{westm07a, westm07b}. The variable S/N nature of the observations makes it difficult to determine whether at larger radii we stop detecting this broad component because it actually does not exist, or whether as the H$\alpha$ emission becomes weaker with increasing galactocentric distance, it fades into the continuum noise. Detailed inspection of the line profiles show that we can identify the broad component within a definable region (hereafter referred to as the `broad component region'), $70\times 50$~arcsecs ($\sim$$700\times 500$~pc), approximately centred on SSC A and roughly aligned with the galaxy major axis. This region is outlined with a solid line in Fig.~\ref{fig:finder}.

The fact that this broad component region has an extent approximately equivalent to the H$\alpha$-bright disc could suggest that the broad emission simply fades into the continuum at this point and becomes undetectable. However we find evidence supporting a definite physical extent to the existence of the broad component. Fig.~\ref{fig:sp_eg_fits}e shows an example H$\alpha$ profile from just outside this broad component region (at a distance from SSC A of $\approx$ 330~pc), and clearly shows how it can be modelled with two Gaussians of similar narrow widths, with no requirement for a broad underlying component. Correspondingly, we find spaxels in a number of places just within the border of the outlined region with an equivalent S/N as the example shown, but clearly exhibiting the broad-line component. Therefore, we interpret these results to mean that the broad line does physically cease to exist at the boundary of this region. This finding has important consequences that we discuss in Section~\ref{sect:inner}.

Fig.~\ref{fig:sp_vel} shows the spatial variation of the H$\alpha$ radial velocities with respect to the systemic velocity of the galaxy, for which we have adopted the value of $v_{\rm sys} = -80$~\kms\ \citepalias[][and references therein]{westm07a}. In general, the velocity of C1 (left-hand panel) appears consistent with the systemic velocity, and remains fairly constant over the field, varying between $-40$ and +40~\kms\ relative to $v_{\rm sys}$. We note that the C1 velocities in gaps between filaments are generally redshifted compared to the velocities measured in the bright filaments themselves. We do not find evidence for the east-west galactic rotation seen in H\one\ emission. A careful comparison of the particular regions covered by our SparsePak pointings to the H\one\ velocity map of \citet[][their fig.~4]{muhle05}, shows that in fact most of the equivalent H\one\ gas is either consistent with being at $v_{\rm sys}$ or very disturbed with no systematic pattern \citep[see also][]{stil02, heckman95}. It is likely therefore that any signature of rotation in the ionized halo is either masked by the disturbed, strongly disrupted velocity field, or dominated by the effects of large-scale bubble expansion.

\begin{table}
\centering
\caption {Measured and derived supershell properties}
\label{tbl:sp_shells}
\begin{tabular}{llll}
\hline
Supershell & $v_{\rm exp}$ & Diameter & Dynamical \\
& (\kms) & (kpc)$^{a}$ & age (Myr)$^{a}$ \\
\hline
A & $\sim$90 & $\sim$1.1 & $\lesssim$\,10--15 \\
B & $\sim$85 & $\sim$1.1 & $\lesssim$\,10--15 \\
E & $\lesssim$\,50 & $\sim$1.3 & $\lesssim$\,25 \\
F & $\lesssim$\,100 & $\sim$1.3 & $\lesssim$\,25 \\
\hline
\end{tabular}

$^{a}$ see Section~\ref{sect:outer}.
\end{table}

The C2 radial velocity map (Fig.~\ref{fig:sp_vel}, right panel) shows the velocity variation of the second Gaussian component. As described in Section~\ref{sect:line_fitting}, this is by definition the fainter of the two brightest components fitted, and in all cases either represents a broad underlying component or the other half of a split-line profile, depending on what was required to fit the integrated line shape. Within the central $\sim$500~pc, inside the broad component region, the velocities measured are very close to those found for C1. However at larger radii, where C2 is the fainter component of a split-line, we see evidence for ordered galactic-scale outflows. In the north, the blueshifted component is stronger (therefore in our convention assigned to C1), and has velocities ranging between $-20$ and 0~\kms\ (relative to $v_{\rm sys}$), whereas the redshifted component (C2) has relative velocities of up to +150~\kms. The maximum expansion velocity ($\sim$85~\kms; calculated as half the difference between the radial velocities of the two individual components) is found to the north-west of the polar axis corresponding to the centre of shell B (see Fig.~\ref{fig:shells}), and agrees well with the measurements of \citet{martin98}. In the north-east, a small number of split-line detections are coincident with the position of shell E (Fig.~\ref{fig:shells}), implying expansion velocities of up to 50~\kms. In the south, the redshifted component is stronger (hence assigned to C1) and exhibits velocities between 0 and +20~\kms. The blueshifted component (C2) has corresponding radial velocities of up to $-160$~\kms\ (relative to $v_{\rm sys}$), giving a peak expansion velocity of 90~\kms\ to the south-west of SSC A, coincident with the centre of shell A \citep[Fig.~\ref{fig:shells}; also agreeing well with the velocities quoted in][]{martin98}. Expansion velocities of $\gtrsim$40~\kms\ are observed throughout the whole length of the cavity encompassed by shell A and the western arm. Blueshifted C2 velocities of $-50$ to $-75$~\kms\ are also seen in the northern half of position 3(F), implying expansion velocities of up to 100~\kms\ for shell F. We summarise the expansion velocity measurements for the four supershells in Table~\ref{tbl:sp_shells}, together with their diameters and dynamical ages derived in Section~\ref{sect:outer}.

\begin{figure*}
\centering
\includegraphics[width=0.8\textwidth]{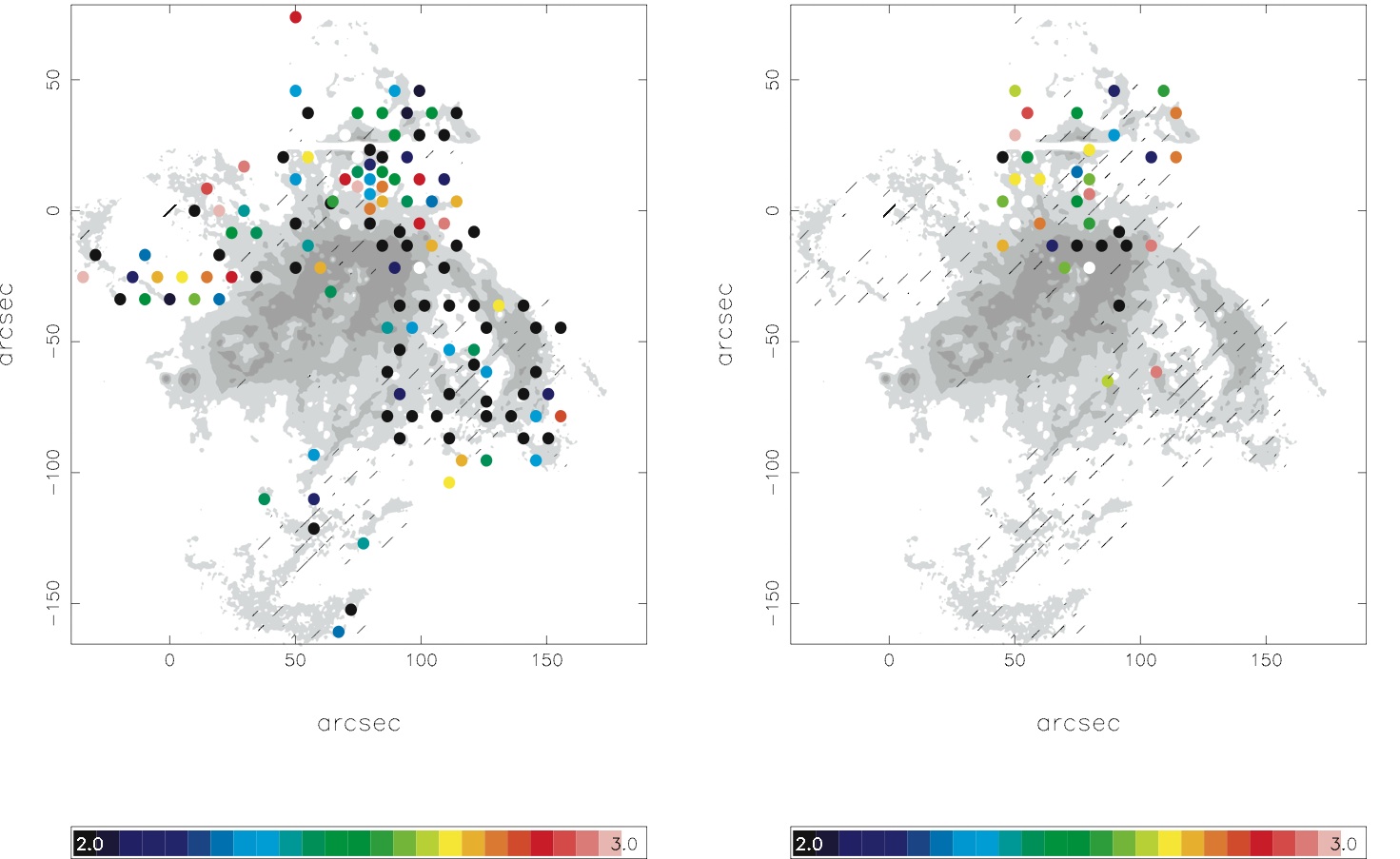}
\caption{Spatial distribution of electron density (in units of log \cmt) for the four SparsePak fields for \emph{left:} C1 and \emph{right:} C2, calculated from the flux ratio of the [S\two] doublet. The colour map used is shown beneath each plot and ranges upwards from the low density limit of this indicator \citep[$\sim$100~\cmt;][]{osterbrock89}.}
\label{fig:sp_elecdens}
\end{figure*}

In three spaxels in position 2(B) (numbers 64, 73 and 76), coinciding with the centre and western edge of shell B, we detect three distinct H$\alpha$ line components similar to the example shown in Fig.~\ref{fig:sp_eg_fits}c. The `extra' component, with respect to the neighbouring double-component (narrow + broad) lines, lies at a velocity close to $v_{\rm sys}$, so could represent quiescent halo gas or be associated with part of an as yet undetected shell or bubble.

The consistent reversal of strengths of the two major components of the split line from the north to south agrees well with previous studies \citep{heckman95, martin98}, and demonstrates there is a definite preferred outflow direction in both the northern and the southern hemispheres. This strengthens the hypothesis that the actual axis of the NGC 1569 galactic outflow may be inclined to the observer \citep{stil02}, in the sense that the northern outflow is tipped away from our sightline, rather than simply that the bubbles are expanding out of the disc along random paths of least resistance.

\subsection{Electron Density}
Fig.~\ref{fig:sp_elecdens} shows a map of the electron density, \Ne, as derived from the flux ratio of the [S\two]$\lambda\lambda$6717,6731 doublet, assuming $T_{\rm e}=10^{4}$~K. [S\two] emission is detected in all spaxels that also have bright H$\alpha$ emission, but often the densities implied were below the low density limit for this indicator\citep[$\sim$100~\cmt;][]{osterbrock89}. Hence, we have set the minimum value of the colour range to 100~\cmt, meaning that spaxels where [S\two] emission was detected but imply densities below the low density limit are shown as black. Spaxels where [S\two] is not detected or where an unphysical flux ratio is measured (resulting easily from the large measurement errors) are represented by hatched lines.

\begin{figure*}
\centering
\begin{minipage}{7cm}
\includegraphics[width=6.8cm]{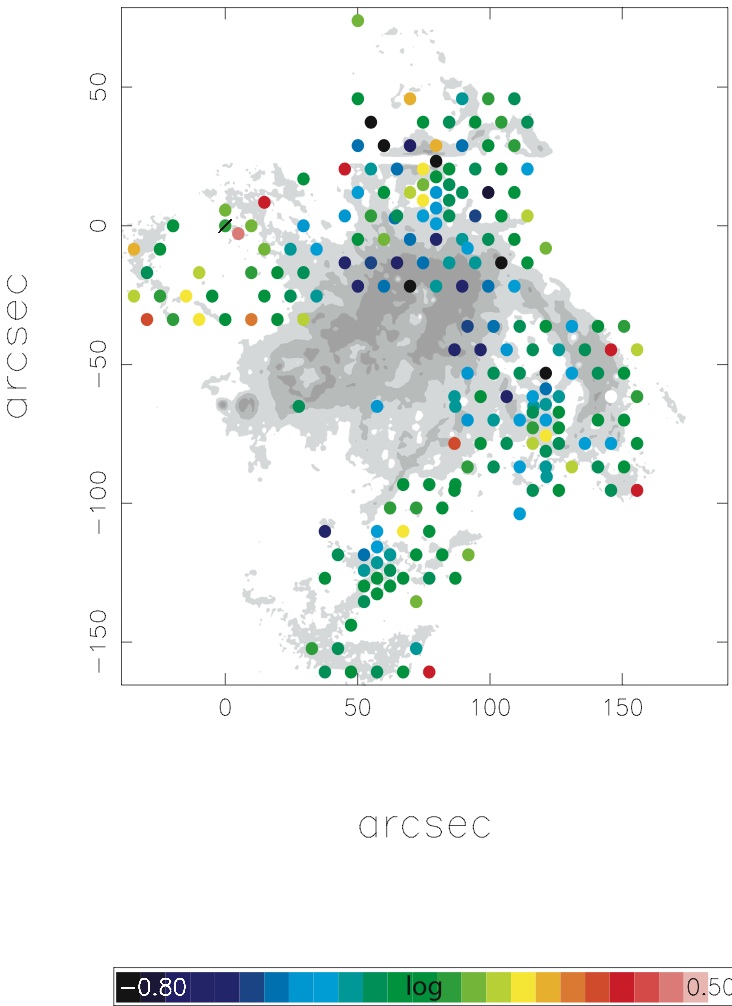}
\end{minipage}
\hspace{0.3cm}
\begin{minipage}{7cm}
\includegraphics[width=6.8cm]{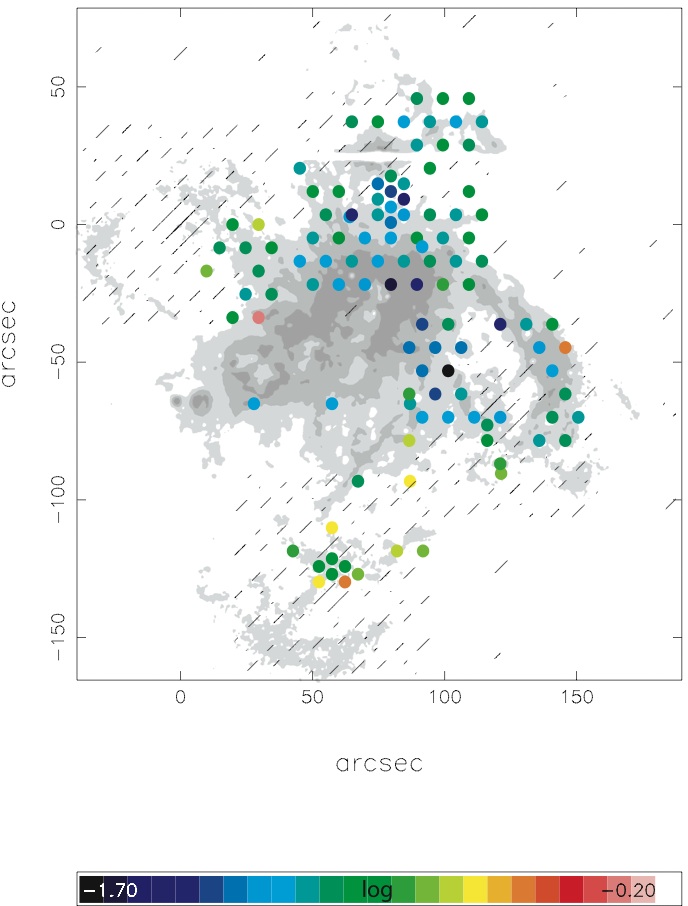}
\end{minipage}
\caption{\emph{Left:} the spatial distribution of the [S\two]($\lambda$6717+$\lambda$6731)/H$\alpha$ ratio and \emph{right:} [N\two]$\lambda$6583/H$\alpha$ ratio for the four SparsePak fields (C1 only).}
\label{fig:sp_sii_ha}
\end{figure*}

The majority of the ionized gas in the outer-wind region is at or below the low density limit, as expected for a rarefied expanding outflow. Some regions in shell B [in position 2(B)] and shell E (in position 1(E); see Figs~\ref{fig:shells} and \ref{fig:finder}) exhibit measurable densities in C1 (left-hand panel) of order 500~\cmt. In parts of position 2(B), and a few spaxels in position 4(A), we are able to detect two line components and measure the electron densities of both to be of the order $\sim$$500\pm 300$~\cmt. Unfortunately, the multi-component nature of the lines and the variable S/N have meant that the associated uncertainties on the measurements are high.

\subsection{Nebular diagnostics}

The flux ratios of [S\two]/H$\alpha$ and [N\two]/H$\alpha$ can be used as an indicators of the excitation level of the gas, and can be used to ascertain whether the gas is shock-excited, since the forbidden-line emission (particularly [S\two]) is enhanced in shocks \citep{dopita95, calzetti04}. Unfortunately, the strength of [N\two] in NGC 1569 is very weak compared to H$\alpha$ \citep[of order 2--5 per cent; see also][]{heckman95} due to the low metallicity of the gas, but nevertheless is detected in the brighter regions.

Fig.~\ref{fig:sp_sii_ha} shows spatial maps of the [S\two]($\lambda$6717+$\lambda$6731)/H$\alpha$ and [N\two]$\lambda$6583/H$\alpha$ ratios. The limiting factor in both these maps is the S/N of the fainter forbidden-line emission; only where we have been able to detect emission have we been able to measure the line ratios. Having said that, the S/N of the forbidden lines is high enough in some places for multiple components to be detected, so to avoid confusion we use only the ratio of the brightest component (C1) identified by our Gaussian fitting routine for this ratio.

The majority of spaxels where [S\two] emission is detected have a log([S\two]/H$\alpha$) $\approx$ $-0.5$ to 0, and the highest ratios are generally associated with the faintest H$\alpha$ emission whereas the lowest ratios are found where the H$\alpha$ flux is highest. Spaxels with detectable [N\two] emission have a log([N\two]/H$\alpha$) $\approx$ $-0.2$ to $-1.3$, and the lowest ratios are associated with the brightest H$\alpha$-emitting gas. As with [S\two]/H$\alpha$, this trend is correlated well with galactocentric distance, where the two ratios both increase from the region around SSC A towards the outer halo.

These trends agree well with the measurements of \citet{heckman95}, who mapped the [S\two]/H$\alpha$ and [N\two]/H$\alpha$ ratios out to $\sim$$\pm$$50''$ from SSC A using two perpendicular slits, except that they found values of log([S\two]/H$\alpha$) $\approx$ $-1.2$ at the location of SSC A. However, \citeauthor{heckman95}~did not mention whether or not they fitted the [S\two] lines with multiple components, so we do not know whether they measured the flux ratio of the integrated line-flux, or that of just the brightest component. If the former was true, then a significant part of the discrepancy could arise from the varying contribution from the different line components. Furthermore, most of the central region covered by their slits is not well sampled by our SparsePak fibres, and we may easily have missed the very low ratios they find in the inner $20''$ or so.

\begin{figure}
\centering
\includegraphics[width=0.48\textwidth]{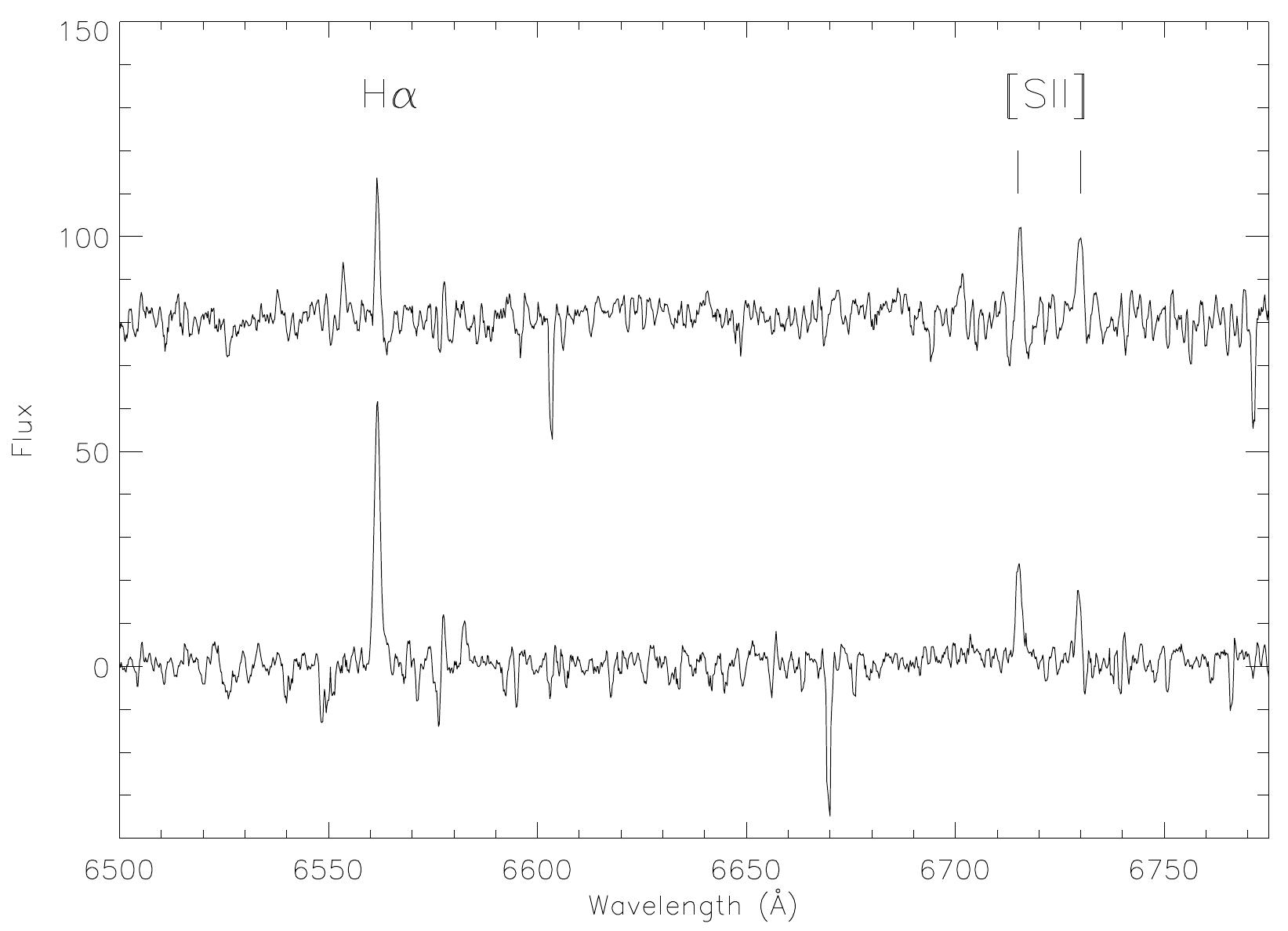}
\caption{Two examples of spectra that show high [S\two]/H$\alpha$ ratios. The lower spectrum is from the centre of position 3(F) (fibre 25; log([S\two]/H$\alpha$) $\approx$ $-0.25$), and the upper spectrum (offset by 80 flux units) is from the far south-east of position 1(E) (fibre 12; log([S\two]/H$\alpha$) $\approx$ 0.20). Both are smoothed by 5~\AA{}, and shown on an arbitrary but relative flux scale.}
\label{fig:eg_spec_ratio}
\end{figure}

A number of spaxels exhibit line ratios that are worth drawing attention to. Fig.~\ref{fig:eg_spec_ratio} shows two such spectra from the outer-wind regions with simultaneously very high [S\two]/H$\alpha$ and very low [N\two]/H$\alpha$ ratios. The upper spectrum has a [S\two]/H$\alpha$ ratio of $\sim$1.25 (log([S\two]/H$\alpha$) $\sim$ 0.10), representative of the whole south-east of position 1(E) where most of the highest [S\two]/H$\alpha$ ratios are found. Since there is no detectable [N\two] emission (see Fig.~\ref{fig:sp_sii_ha}, right panel), we can place an upper limit on the [N\two]/H$\alpha$ ratio of $<$0.1. The lower spectrum shows a similar situation from the centre of position 3(F) (arc 1) where [S\two]/H$\alpha$ is $>$0.5 and [N\two]/H$\alpha$ is $<$0.1. High [S\two]/H$\alpha$ ratios are also found clustered in the centre of position 2(B), position 3(F) (arc 1; lower spectrum in Fig.~\ref{fig:eg_spec_ratio}), and position 4(A).

A ratio of log([S\two]/H$\alpha$) $>$ 0 lies in the shocked region of a [O\three]/H$\beta$ vs.~[S\two]/H$\alpha$ diagram, for any sensible values of [O\three]/H$\beta$ \citep{dopita95, calzetti04}, thus indicating that at least shell E may be the result from an outward starburst-driven shock. Shocks are certainly expected in the outflow, since the hot superbubble will be expanding into preexisting H\one{} gas and/or other halo material, which is effectively at rest with respect to the outflow. The existence of shocks is also supported by the detection of strong X-ray emission from these shells \citep[particularly shell E where the corresponding H$\alpha$ emission is weak;][Section~\ref{sect:outer}]{martin02}. The shock excited component may become more obvious with increasing radius due to the fact that further from the central star clusters, the excitation of the gas is less dominated by the effects of photoionization.

\section{Comparison between SparsePak and Gemini GMOS/IFU data} \label{sect:comparison}

Before we discuss the implications of the broad component and what it may represent physically, we should compare its characteristics to the corresponding broad component found using our Gemini GMOS/IFU observations \citepalias{westm07a, westm07b}.

One way to compare the datasets quantitatively is by examining the flux ratio of C2/C1 -- an important indicator of the significance of the two components\footnote{We must be careful in discussing this component since here 'C2' refers to both a broad underlying component and the other half of a split line profile, depending on context, whereas in \citetalias{westm07a} and \citetalias{westm07b} 'C2' refers only to the broad underlying component.}. Although the quality of the data presented here is not high enough to identify any trends in this ratio across the observed fields, what does deserve attention is why we find such a high average value of C2/C1 ($\sim$0.7) compared to the value found from the GMOS data \citepalias[$<$0.3;][]{westm07b}.

It is simple to understand a high C2/C1 ratio for cases where C2 is one half of a split-line profile: in our assignment convention C1 is by definition the brighter of the two components (C2/C1 $<$ 1), and since in general the intensity of the two split-line components is comparable (to within a factor of a few), high C2/C1 ratios are expected. However, we also find C2/C1 ratios of $\sim$0.2--0.7 within the broad component region, where from our GMOS data we find C2/C1 $<$ 0.3.

\begin{figure*}
\centering
\begin{minipage}{7cm}
\includegraphics[width=7cm]{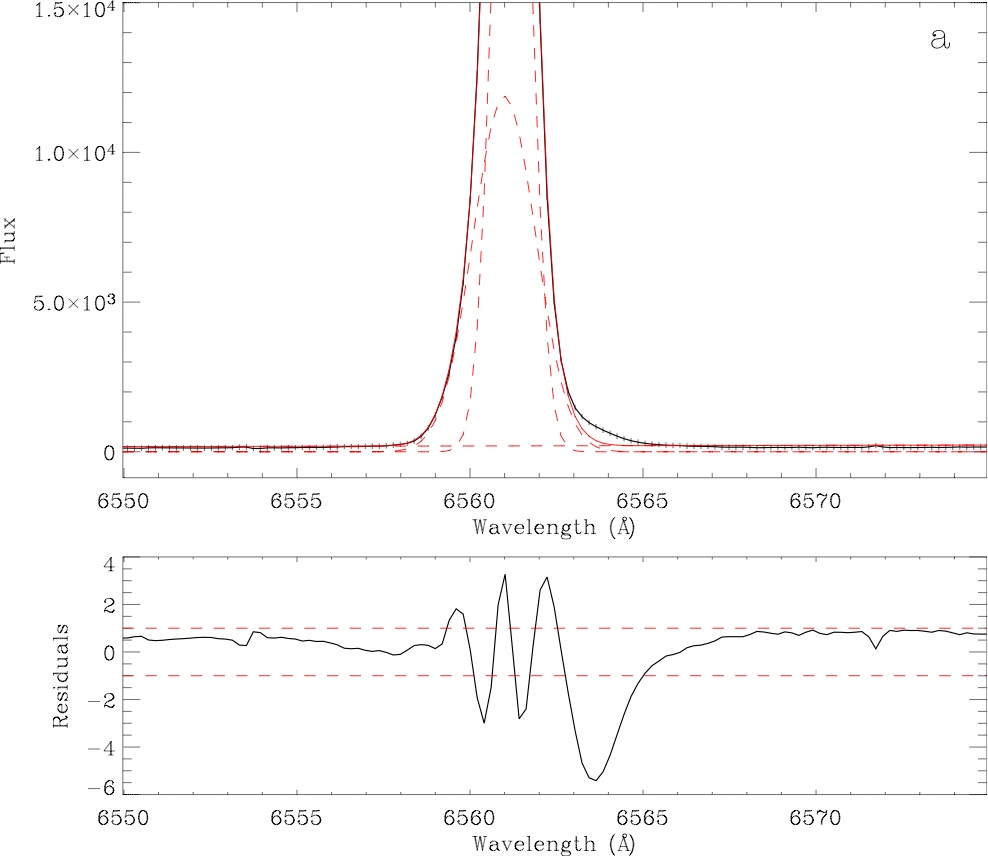}
\end{minipage}
\hspace{0.3cm}
\begin{minipage}{7cm}
\includegraphics[width=7cm]{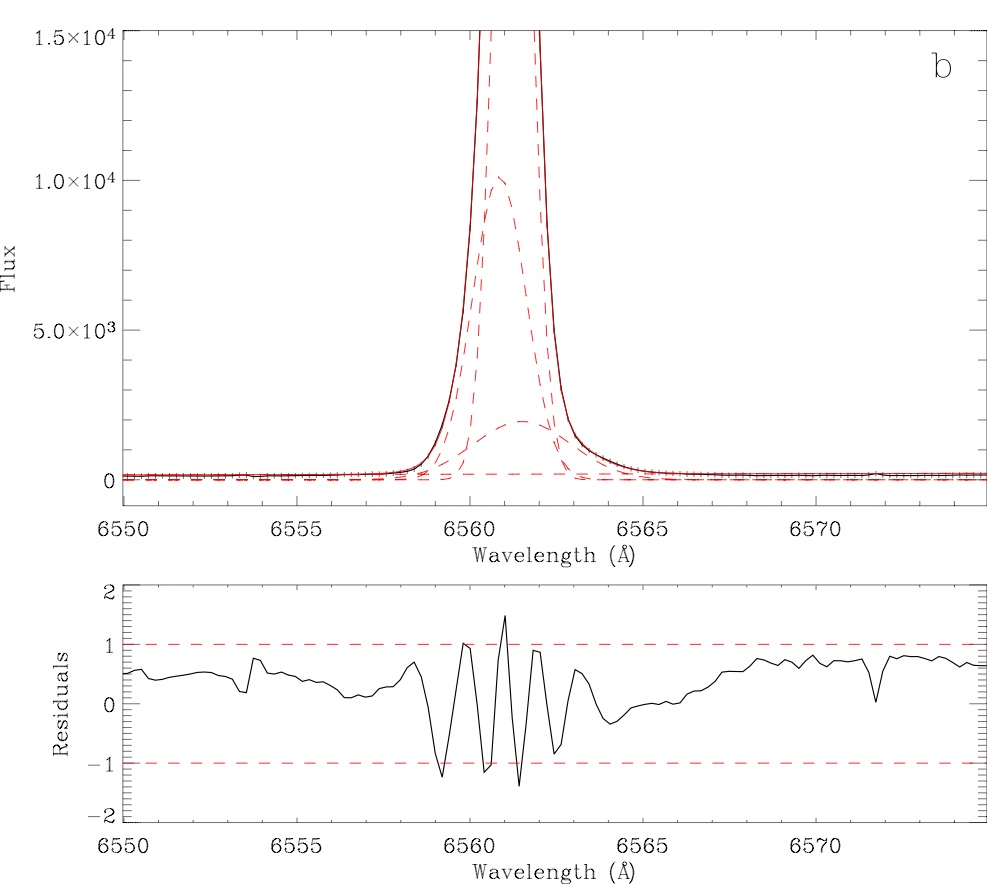}
\end{minipage}
\vspace{0.2cm}
\caption{H$\alpha$ profile from fibre 50 of SparsePak position 2(B) \citepalias[nearest to GMOS IFU position 2 of][see Fig.~\ref{fig:finder}]{westm07a}, illustrating how a more careful modelling of the line profile reveals that a three-component fit (panel b) characterises the integrated shape more accurately than the automatically determined double-Gaussian fit (panel a). The addition of a third component and the correct assignment of the profiles results in the FWHM of C2 increasing from 100 to 153~\kms{} and the C2/C1 ratio changing from 0.69 to 0.17, now in much better agreement with the GMOS data for the same region \citepalias{westm07a}. Observed data is shown by a solid black line, individual Gaussian fits by dashed red lines (including the straight line continuum level fit), and the summed model profile in solid red. Flux units are arbitrary but on the same scale. The fit residuals are shown below the spectrum in both cases.}
\label{fig:sp_c2_assignment}
\end{figure*}

To investigate this discrepancy, we re-examined a number of H$\alpha$ profiles in the SparsePak spaxels closest to SSC A and the GMOS IFU fields (see inset to Fig.~\ref{fig:finder}), and found that in some of these cases a subtle extra component had been overlooked by the automatic fitting procedures. Fig.~\ref{fig:sp_c2_assignment} shows the H$\alpha$ profile from fibre 50 of SparsePak position 2(B) \citep[the fibre nearest to GMOS IFU position 2 and covering part of the bright H\two\ region No.\ 2;][]{waller91}, and is the most obvious example of how, after close inspection, a three-component fit is found to be more appropriate (shown in panel b) than the original, automatically determined double-Gaussian fit (panel a). The residuals shown below each plot clearly demonstrate the improvement in fit quality, even though the formal $\chi^{2}$ improvement is not statistically significant. In this particular case, the effect of changing C2 of the double-Gaussian fit to C2 of the triple-Gaussian fit (now correctly assigned to a broad underlying component rather than a second narrow component) is to increase its FWHM by 50~\kms{} to $\sim$150~\kms, and to decrease the C2/C1 flux ratio from 0.69 to 0.17, now in much better agreement with that found from the GMOS data. This mis-fitting also offers an explanation for why the widths of C2 near SSC A measured from our SparsePak data are much less than the average values measured from the GMOS observations \citepalias[200--300~\kms][]{westm07b}. 

Unfortunately, this kind of detailed fitting can only be done for the highest S/N spectra found near the centre of NGC 1569, so cannot be used reliably to track the C2/C1 ratio over the full extent of the broad component region.

\section{Discussion} \label{sect:disc}
The central part of the ISM of NGC 1569 has been severely disturbed by the effects of an intense starburst event. Through deep H$\alpha$ imaging and wide-field spatially resolved spectroscopy, we have investigated the state of the ionized gas in the halo of this galaxy. A detailed decomposition of the H$\alpha$ line profiles shows that within a distinct region $\sim$$700\times 500$~pc in size and roughly centred on the location of SSC A, they are composed of a bright, narrow component with an underlying, broad component (FWHM $\lesssim$ 150~\kms). We refer to this region as the broad component region. At larger radii, the broad component disappears and the profile shape becomes double-peaked, exhibiting characteristics of shell expansion. 

\subsection{The inner galaxy and the broad component} \label{sect:inner}
In \citetalias{westm07a}, we discussed in detail the possible mechanisms that could give rise to the underlying broad component found in the central regions of the galaxy. We concluded that the most likely cause is the impact of strong winds from the massive star clusters setting up turbulent mixing layers \citep{begelman90} on the surfaces of ISM cloud clumps. Models predict that the ionized gas emission from these layers would exhibit line widths characteristic of the turbulent velocities within the layers \citep{slavin93}. The subsequent ablation and/or evaporation of material from these surfaces results in a pervasive, highly turbulent velocity field, explaining why the broad underlying emission is present throughout the inner regions of the burgeoning outflow.

Through the arguments presented in Section~\ref{sect:comparison}, we have established that the properties of the broad line component seen with SparsePak in the regions near to SSC A are equivalent to the broad line found in our Gemini GMOS/IFU observations \citepalias{westm07a, westm07b}. We can therefore confidently associate the two broad line components with each other and thus the physical mechanisms that produce them. However, the low spatial-resolution of the SparsePak array introduces a greater uncertainty to the degree unresolved kinematical components contribute to the line width. For example, one SparsePak fibre covers a similar area to the whole field of view (FoV) of the GMOS IFU ($5\times 3.5$~arcsecs; see Fig.~\ref{fig:finder}), and in \citetalias{westm07b} we found C2 to have a radial velocity spread of $>$90~\kms\ over just one IFU field. We must therefore accept that all our FWHM measurements will be artificially elevated by unresolved components at the level of a few tens of \kms.

An interesting and pertinent question arises from our study: why might the extent of the broad underlying emission be limited to a specific area (the broad component region), and what significance does this have? Our data show that the radius at which the broad line is no longer identifiable also roughly corresponds to the point that the H$\alpha$ profile starts showing a secondary narrow kinematic component. We speculate therefore that this radius marks a significant transition point in the development of the galactic outflow, where ordered expansion begins to dominate over turbulent motion, as is expected as a superbubble emerges from a disc. This possibly significant result warrants detailed further investigation at higher spatial-resolution and S/N to better quantify this transition region. 

\subsection{The ionized halo} \label{sect:outer}

\begin{figure*}
\begin{center}
\begin{minipage}{12cm}
\includegraphics[width=12cm]{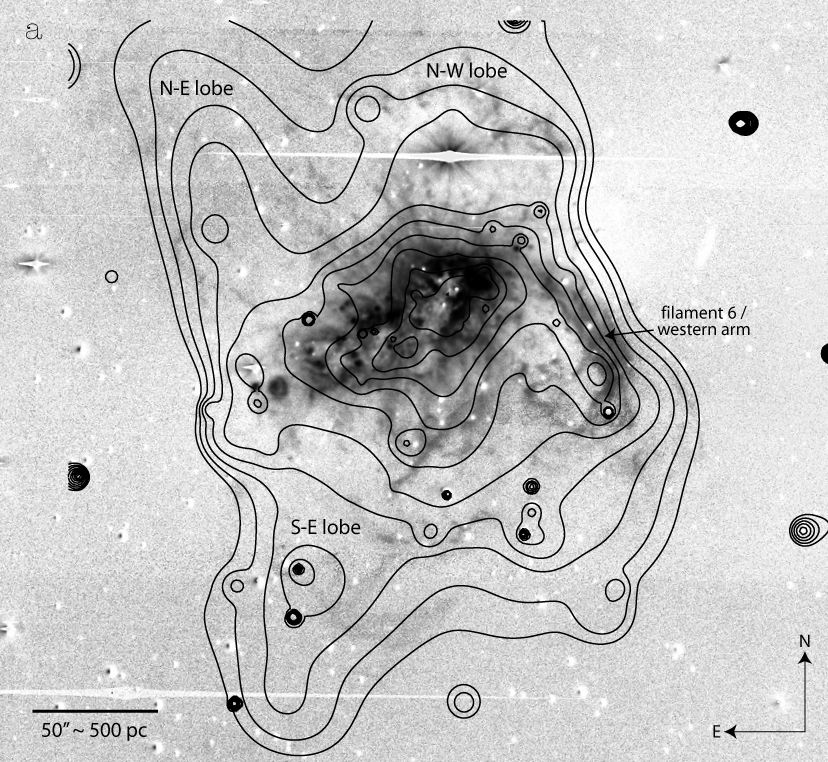}
\end{minipage}
\begin{minipage}{12cm}
\vspace{0.3cm}
\includegraphics[width=12cm]{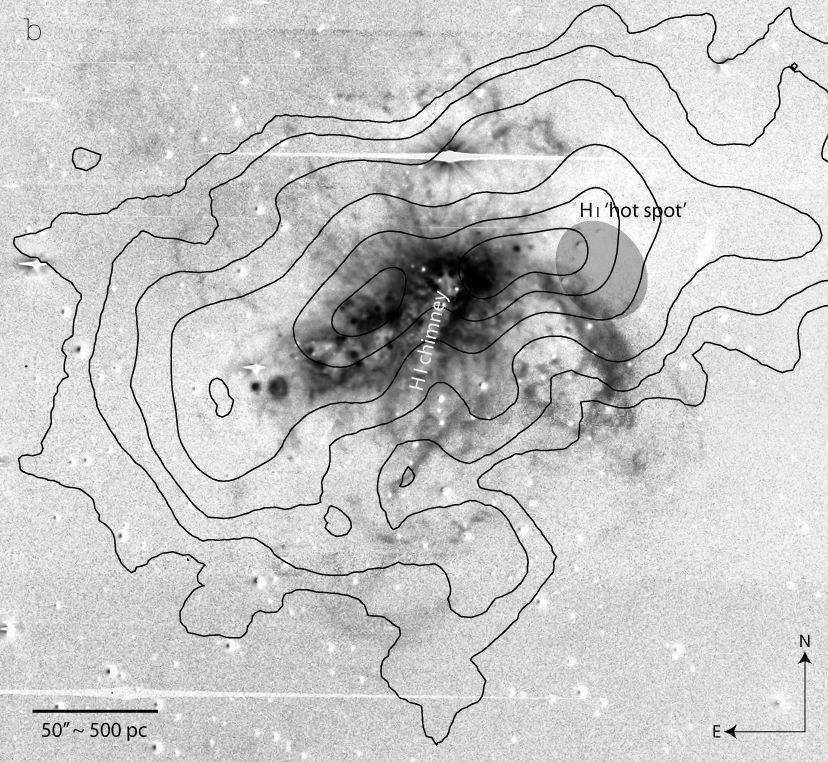}
\end{minipage}
\end{center}
\caption{(a) \textit{Chandra} medium-band X-ray contours \citep{martin02} and (b) H\one\ column density contours \citep{muhle05} overlaid on the WIYN H$\alpha$ image of Fig.~\ref{fig:wiynha}. The shaded area on the bottom panel indicates the H\one\ `hot-spot' (region of velocity crowding); a number of other features discussed in the text are also labelled.}
\label{fig:xray_HI}
\end{figure*}

The ionized gas morphology suggests a situation whereby individual bubbles, blown by a star cluster or SNe, expand into the halo through channels or `chimneys' in the ISM distribution that were carved by previous generations of bubbles. Once a channel is cleared, swept up material can flow through the same channel with much less energy losses, i.e.~less mechanical energy is converted into thermal energy. In this interpretation, the outflow could be likened to a series of cannon bursts, where each explosive event (possibly the formation of a massive star cluster or a collection of star clusters) has produced one of the bubbles we observe today.

In Fig.~\ref{fig:xray_HI}, we overplot contours representing medium-band X-ray emission (panel a) and H\one\ column density (panel b) on the H$\alpha$ image of Fig.~\ref{fig:wiynha}. It is striking how well the X-ray emission morphology follows the H$\alpha$ distribution \citep[as noted by previous authors;][]{heckman95, martin02}. However correspondences can also be found in the H\one\ map.

Firstly, the H$\alpha$ counterpart to the H\one\ chimney \citep{israel90, muhle05} can be clearly seen extending to the south-east of SSC A (labelled on Fig.~\ref{fig:xray_HI}b). This chimney appears to be the primary channel through which a significant amount of the disc ISM is escaping and forming superbubble F. Less prominent `holes' in the H\one\ distribution also correspond well to the centres of shells A and E. Secondly, \citet{muhle05} identify a `hot spot', a region of velocity crowding, on the western edge of the disc that they interpret as resulting from the impact of infalling H\one\ gas from an interacting cloud. This area, shaded on Fig.~\ref{fig:xray_HI}b, is just exterior to the peak of the H$\alpha$ emission and the location of the large active H\two\ complex 2 \citep{waller91}. Multiwavelength observations of this part of the disc support the existence of a large molecular gas reservoir \citep{taylor99} and the existence of continuing star- and cluster-formation (\citealt{tokura06}; \citetalias{westm07a}), suggesting a possible reason for the irregular, off-centre location of the starburst and the two SSCs A and B.

Perhaps more significantly, however, the hot-spot is located just to the west of filament 6 (the bright western arm, shell A), where a strong X-ray enhancement is also detected \citep[Fig.~\ref{fig:xray_HI}a;][]{martin02}. Why the western arm is so bright and part of such a large, coherent structure has been a standing problem since it was first detected \citep{hodge74}. We suggest that shell A was initially formed in a similar fashion to the other superbubbles, but as it expanded out into the halo it encountered a barrier formed by the infalling cold gas. The thermalisation of the outflowing material by the interaction resulted in increased radiative losses (showing up in both H$\alpha$ and X-rays), hence the slowing of the bubble's expansion on the west side. The bubble then became distorted as the outflow was diverted from its original path, producing the structure that we see today.

This leads us to consider the fate of the bubbles and the outflow itself. \citet{martin98} interpreted the fact that the velocity separation of the two line components in shell A were still increasing when the fainter side faded below her detection limit as evidence that this bubble may have ruptured. However, our deeper SparsePak observations show that this is not the case. We measure decreasing velocity separations in both shells A and B at the very edge of our combined FoV. In shell A, the expansion velocity peaks $\sim$330~pc south of SSC A (at 90~\kms), then begins to decrease until we can no longer detect H$\alpha$ emission. In shell B, the expansion velocity peaks $\sim$375~pc north of SSC A (at 85~\kms), we again see a decrease in $v_{\rm exp}$ up to the edge of our FoV at distances of $>$1~kpc (although the S/N of our spectra at these radii is quite low). In support of this, signatures of shell blow-out are not seen in the H$\alpha$, H\one\ \citep{muhle05} or X-ray \citep{martin02} morphologies (Figs~\ref{fig:wiynha} and \ref{fig:xray_HI}). Firstly, there are no large-scale holes or under-densities in the H\one\ column density of the halo that could be indicative of outflow channels. Secondly, the soft- and medium-band X-ray emission morphologies are very well matched to the H$\alpha$ distribution, indicating that hot gas is not venting out of the bubble interiors \citep[Fig.~\ref{fig:xray_HI}a;][]{martin02}.

These conclusions do not, however, exclude the possibility of future blow-outs. The best estimates for the escape speed in NGC 1569 are of the order of $\sim$100~\kms\ \citep{martin98}, which when compared to our H$\alpha$ expansion velocity measurements, or to the kinetic temperature of the X-ray emitting gas, imply that the material in the shells may have the ability to escape the galaxy's gravitational potential eventually.

Measuring sizes of the four superbubbles from Fig.~\ref{fig:shells}, and using the expansion velocities calculated in Section~\ref{sect:maps}, we can estimate approximate dynamical ages for the shells. Shells A and B have radial extents of $\sim$1.1~kpc and current expansion velocities on the order of 80--90~\kms, giving dynamical ages of $\lesssim$10--15~Myr. Estimating the true extents of superbubbles E and F is difficult considering their low surface brightnesses and more chaotic morphologies. Using the dashed lines plotted on Fig.~\ref{fig:shells} as a guide, we measure diameters for these bubbles of $\sim$1.3~kpc, and assuming constant expansion velocities of $\sim$50~\kms, this gives ages of $\sim$25~Myr (or less if the shells were expanding faster in the past). These measurements are summarised in Table~\ref{tbl:sp_shells}.

We can compare these ages to the star and cluster formation histories derived in previous work. We know NGC 1569 to have undergone a burst of $\sim$constant rate star formation that stopped $\sim$5--10~Myr ago from resolved stellar population modelling \citep{greggio98}. More specifically however, we know from evolutionary synthesis modelling that SSCs A and B were formed $\sim$10--30~Myr ago, and cluster 30 between $\sim$30--100~Myr ago \citep{anders04}. \citeauthor{anders04} also concluded that the last major burst of cluster formation must have occurred some 25~Myr ago. These results fit in very well with our derived bubble ages, and lend support to the suggestion that each bubble forms after the creation of, for example, a massive star cluster.

\section{Summary and conclusions} \label{sect:concs}
We have presented WIYN MiniMo deep H$\alpha$ imaging of NGC 1569 covering a field-of-view of $9.5\times 10.7$~arcmins. The depth and large dynamic range of the observations have enabled the identification of previously undetected faint ionized filaments in the halo and the study of the gas morphology right down into the bright, central regions of the disc. We have also presented WIYN SparsePak ``formatted field unit'' observations covering the outer galactic wind flow of NGC 1569 in four pointings with integration times of 4--4.5 hours per field. The large diameter of the SparsePak fibres makes this instrument ideal for probing the faint ionized gas found in the halos of galaxies. This light-collecting power allowed us to choose a high-resolution spectrograph set-up, enabling us to characterise the line profile shapes of the important nebular diagnostic lines of H$\alpha$ and [S\two] to an accuracy limited only by the S/N achieved. We now summarise our main findings.

\begin{itemize}
  \item We find H$\alpha$ emission out to radii of $\sim$1.5~kpc from the disc, and detect emission in almost every SparsePak fibre over the combined FoV. The presence of such an extensive system of ionized filaments results from the ongoing starburst that is supplying both mechanical energy to eject material and the Lyman continuum luminosity to keep it photoionized.
  \item Through detailed Gaussian line fitting, we find that within a distinct region $\sim$$700\times 500$~pc in size, roughly centred on the location of SSC A, the nebular emission line profiles are composed of a bright, narrow (FWHM $\lesssim$ 70~\kms) component with an underlying, broad component (FWHM $\sim$ 150~\kms). At larger radii, we find two narrow components to the H$\alpha$ line, each representing one half of a split-line profile.
  \item By comparing our results to observations of the central regions directly surrounding SSC A \citepalias{westm07a, westm07b}, we conclude that the physical mechanisms that give rise to the underlying broad emission seen within this zone must be the same as within the regions directly surrounding SSC A sampled by Gemini GMOS/IFU observations. The broad emission is most likely to result from turbulent mixing layers on the surface of cool gas clumps set up by the impact of the hot, fast-flowing cluster winds, and from evaporation and/or ablation of material from the clumps.
  \item The extent of this broad component region, coincident with the point at which we start observing signatures of large-scale bubble expansion, may indicate a transition point where ordered expansion begins to dominate over turbulent motion. Further observations are needed to investigate this in more detail.
  \item By combining our deep H$\alpha$ imaging and spectroscopy, we redefine the spatial extents of superbubbles A and B, and confirm their published \citep{martin98} expansion velocities (90 and 85~\kms, respectively). We estimate the dynamical ages of these bubbles to be $\sim$10--15~Myr. Contrary to what has been previously suggested \citep{martin98}, we find no kinematic or morphological evidence to suggest that either of these two superbubbles have ruptured and are venting their interiors into the galactic halo.
  \item Our data indicate that the halo of NGC 1569 contains only 4 superbubbles. Following the terminology introduced by \citet{martin98}, we propose that her superbubble F should should encompass the whole south-eastern bubble complex, where the velocity ellipses identified by \citet{martin98} and used to define shells G and F, are simply one level of a `hierarchy of structure'. Furthermore, we propose that superbubble E should encompass what were previously referred to as shells D and E and the large north-eastern X-ray spur \citep{martin02}.
  \item We derive new measurements of the expansion velocity, $v_{\rm exp}$ (calculated from the difference in the radial velocities between the two split-line components), for the superbubble complexes E and F of $v_{\rm exp}\sim50$~\kms{} and $\lesssim$100~\kms, respectively. Assuming a diameter of $\sim$1.3~kpc for these two structures implies dynamical ages of $\lesssim$25~Myr.
  \item The derived ages of the supershells are consistent with the recent cluster formation history of NGC 1569 \citep{anders04}, implying that each shell is associated with a specific star-forming event, such as a young massive star cluster that can provide a large mechanical luminosity from its many type II supernovae.
  \item The consistent reversal of strengths between the blue and red components in the northern and southern outflows provides evidence of preferred outflow directions approximately perpendicular to the inclined and flattened H\one\ disc \citep{stil02}.
  \item In addition to characterising the H$\alpha$ line profile, we have also measured [S\two] derived electron densities and [S\two]$\lambda$6717+$\lambda$6731/H$\alpha$ line ratios. We find that much of the ionized gas in the broad component region and in the outer-wind regions is at or below the low density limit, as is expected for a rarefied outflow. We find the highest [S\two]/H$\alpha$ ratios are associated with the faintest H$\alpha$ fluxes and the largest galactocentric distances. log([S\two]/H$\alpha$) ratios $>$0 are found in the superbubbles E, B and A, indicating that in these regions the gas emission may be significantly shock-excited.
\end{itemize}

In summary, the outflow in NGC~1569 appears to consist of several superbubbles in various phases of development, as noted by \citet{martin98}. This situation is further complicated by the disturbed state of the H\one\ that includes features well out of the main plane of the galaxy (Fig.~\ref{fig:xray_HI}b). Our data confirm this model and thus indicate that the evolution of the outflow will be determined by the development of the superbubbles. In superbubbles A, B, and F ionized gas arcs seen in H$\alpha$ and the X-ray morphology suggest that much of the hot gas still is confined, albeit moving at velocities that are comparable to those needed to escape. Thus the NGC~1569 outflow does not currently appear to be in the form of an approximately steady state galactic wind, even though it may eventually lead to mass loss from the system. It therefore differs from the well known M82 outflow, whose outer regions can be modelled by a supersonic galactic wind \citep[e.g.][]{suchkov94, shopbell98, zirakashvili06}.

\section*{Acknowledgments}
We would 
like to thank Stefanie M\"uhle and Crystal Martin for making their H\one\ and X-ray data available to us. MSW would like to thank the staff at the WIYN Observatory for their support during the observing runs. MSW and LJS thank the University of Wisconsin--Madison for the warm hospitality received in support of this project. JSG's research in this area was partially supported by the University of Wisconsin Graduate School.

\bibliographystyle{mn2e}
\bibliography{/Users/msw/Documents/work/Thesis/thesis/references}

\begin{thebibliography}{}

\bibitem[\protect\citeauthoryear{{Anders}, {de Grijs}, {Fritze-v.~Alvensleben}
  \& {Bissantz}}{{Anders} et~al.}{2004}]{anders04}
{Anders} P.,  {de Grijs} R.,  {Fritze-v.~Alvensleben} U.,    {Bissantz} N.,
  2004, \mnras, 347, 17

\bibitem[\protect\citeauthoryear{{Arp} \& {Sandage}}{{Arp} \&
  {Sandage}}{1985}]{arp85}
{Arp} H.,  {Sandage} A.,  1985, \aj, 90, 1163

\bibitem[\protect\citeauthoryear{{Begelman} \& {Fabian}}{{Begelman} \&
  {Fabian}}{1990}]{begelman90}
{Begelman} M.~C.,  {Fabian} A.~C.,  1990, \mnras, 244, 26P

\bibitem[\protect\citeauthoryear{{Bershady}, {Andersen}, {Harker}, {Ramsey} \&
  {Verheijen}}{{Bershady} et~al.}{2004}]{bershady04}
{Bershady} M.~A.,  {Andersen} D.~R.,  {Harker} J.,  {Ramsey} L.~W.,
  {Verheijen} M.~A.~W.,  2004, \pasp, 116, 565

\bibitem[\protect\citeauthoryear{{Binette}, {Cabrit}, {Raga} \&
  {Cant{\'o}}}{{Binette} et~al.}{1999}]{binette99}
{Binette} L.,  {Cabrit} S.,  {Raga} A.,    {Cant{\'o}} J.,  1999, \aap, 346,
  260

\bibitem[\protect\citeauthoryear{{Calzetti}, {Harris}, {Gallagher}, {Smith},
  {Conselice}, {Homeier} \& {Kewley}}{{Calzetti} et~al.}{2004}]{calzetti04}
{Calzetti} D.,  {Harris} J.,  {Gallagher} J.~S.,  {Smith} D.~A.,  {Conselice}
  C.~J.,  {Homeier} N.,    {Kewley} L.,  2004, \aj, 127, 1405

\bibitem[\protect\citeauthoryear{{de Marchi}, {Clampin}, {Greggio},
  {Leitherer}, {Nota} \& {Tosi}}{{de Marchi} et~al.}{1997}]{demarchi97}
{de Marchi} G.,  {Clampin} M.,  {Greggio} L.,  {Leitherer} C.,  {Nota} A.,
  {Tosi} M.,  1997, \apj, 479, L27

\bibitem[\protect\citeauthoryear{{De Young} \& {Heckman}}{{De Young} \&
  {Heckman}}{1994}]{deyoung94}
{De Young} D.~S.,  {Heckman} T.~M.,  1994, \apj, 431, 598

\bibitem[\protect\citeauthoryear{{Dekel} \& {Silk}}{{Dekel} \&
  {Silk}}{1986}]{dekel86}
{Dekel} A.,  {Silk} J.,  1986, \apj, 303, 39

\bibitem[\protect\citeauthoryear{{Devost}, {Roy} \& {Drissen}}{{Devost}
  et~al.}{1997}]{devost97}
{Devost} D.,  {Roy} J.-R.,    {Drissen} L.,  1997, \apj, 482, 765

\bibitem[\protect\citeauthoryear{{Dimeo}}{{Dimeo}}{2005}]{dimeo}
{Dimeo} R.,  2005, PAN User Guide

\bibitem[\protect\citeauthoryear{{Dopita}, {Kewley}, {Heisler} \&
  {Sutherland}}{{Dopita} et~al.}{2000}]{dopita00}
{Dopita} M.~A.,  {Kewley} L.~J.,  {Heisler} C.~A.,    {Sutherland} R.~S.,
  2000, \apj, 542, 224

\bibitem[\protect\citeauthoryear{{Dopita} \& {Sutherland}}{{Dopita} \&
  {Sutherland}}{1995}]{dopita95}
{Dopita} M.~A.,  {Sutherland} R.~S.,  1995, \apj, 455, 468

\bibitem[\protect\citeauthoryear{{Greggio}, {Tosi}, {Clampin}, {de Marchi},
  {Leitherer}, {Nota} \& {Sirianni}}{{Greggio} et~al.}{1998}]{greggio98}
{Greggio} L.,  {Tosi} M.,  {Clampin} M.,  {de Marchi} G.,  {Leitherer} C.,
  {Nota} A.,    {Sirianni} M.,  1998, \apj, 504, 725

\bibitem[\protect\citeauthoryear{{Heckman}, {Dahlem}, {Lehnert}, {Fabbiano},
  {Gilmore} \& {Waller}}{{Heckman} et~al.}{1995}]{heckman95}
{Heckman} T.~M.,  {Dahlem} M.,  {Lehnert} M.~D.,  {Fabbiano} G.,  {Gilmore} D.,
     {Waller} W.~H.,  1995, \apj, 448, 98

\bibitem[\protect\citeauthoryear{{Hodge}}{{Hodge}}{1974}]{hodge74}
{Hodge} P.~W.,  1974, \apj, 191, L21

\bibitem[\protect\citeauthoryear{{Homeier} \& {Gallagher}}{{Homeier} \&
  {Gallagher}}{1999}]{homeier99}
{Homeier} N.~L.,  {Gallagher} J.~S.,  1999, \apj, 522, 199

\bibitem[\protect\citeauthoryear{{Hunter}, {Hawley} \& {Gallagher}}{{Hunter}
  et~al.}{1993}]{hunter93}
{Hunter} D.~A.,  {Hawley} W.~N.,    {Gallagher} J.~S.,  1993, \aj, 106, 1797

\bibitem[\protect\citeauthoryear{{Hunter}, {O'Connell}, {Gallagher} \&
  {Smecker-Hane}}{{Hunter} et~al.}{2000}]{hunter00}
{Hunter} D.~A.,  {O'Connell} R.~W.,  {Gallagher} J.~S.,    {Smecker-Hane}
  T.~A.,  2000, \aj, 120, 2383

\bibitem[\protect\citeauthoryear{{Israel}}{{Israel}}{1988}]{israel88}
{Israel} F.~P.,  1988, \aap, 194, 24

\bibitem[\protect\citeauthoryear{{Israel} \& {van Driel}}{{Israel} \& {van
  Driel}}{1990}]{israel90}
{Israel} F.~P.,  {van Driel} W.,  1990, \aap, 236, 323

\bibitem[\protect\citeauthoryear{{Izotov}, {Dyak}, {Chaffee}, {Foltz},
  {Kniazev} \& {Lipovetsky}}{{Izotov} et~al.}{1996}]{izotov96}
{Izotov} Y.~I.,  {Dyak} A.~B.,  {Chaffee} F.~H.,  {Foltz} C.~B.,  {Kniazev}
  A.~Y.,    {Lipovetsky} V.~A.,  1996, \apj, 458, 524

\bibitem[\protect\citeauthoryear{{Kobulnicky} \& {Skillman}}{{Kobulnicky} \&
  {Skillman}}{1997}]{kobulnicky97}
{Kobulnicky} H.~A.,  {Skillman} E.~D.,  1997, \apj, 489, 636

\bibitem[\protect\citeauthoryear{{Larson}}{{Larson}}{1974}]{larson74}
{Larson} R.~B.,  1974, \mnras, 169, 229

\bibitem[\protect\citeauthoryear{{Marlowe}, {Heckman}, {Wyse} \&
  {Schommer}}{{Marlowe} et~al.}{1995}]{marlowe95}
{Marlowe} A.~T.,  {Heckman} T.~M.,  {Wyse} R.~F.~G.,    {Schommer} R.,  1995,
  \apj, 438, 563

\bibitem[\protect\citeauthoryear{{Martin}}{{Martin}}{1997}]{martin97}
{Martin} C.~L.,  1997, \apj, 491, 561

\bibitem[\protect\citeauthoryear{{Martin}}{{Martin}}{1998}]{martin98}
{Martin} C.~L.,  1998, \apj, 506, 222

\bibitem[\protect\citeauthoryear{{Martin}, {Kobulnicky} \& {Heckman}}{{Martin}
  et~al.}{2002}]{martin02}
{Martin} C.~L.,  {Kobulnicky} H.~A.,    {Heckman} T.~M.,  2002, \apj, 574, 663

\bibitem[\protect\citeauthoryear{{Melioli}, {de Gouveia dal Pino} \&
  {Raga}}{{Melioli} et~al.}{2005}]{melioli05}
{Melioli} C.,  {de Gouveia dal Pino} E.~M.,    {Raga} A.,  2005, \aap, 443, 495

\bibitem[\protect\citeauthoryear{{Mendez} \& {Esteban}}{{Mendez} \&
  {Esteban}}{1997}]{mendez97}
{Mendez} D.~I.,  {Esteban} C.,  1997, \apj, 488, 652

\bibitem[\protect\citeauthoryear{{M{\"u}hle}, {Klein}, {Wilcots} \&
  {H{\"u}ttemeister}}{{M{\"u}hle} et~al.}{2005}]{muhle05}
{M{\"u}hle} S.,  {Klein} U.,  {Wilcots} E.~M.,    {H{\"u}ttemeister} S.,  2005,
  \aj, 130, 524

\bibitem[\protect\citeauthoryear{{O'Connell}, {Gallagher} \&
  {Hunter}}{{O'Connell} et~al.}{1994}]{oconnell94}
{O'Connell} R.~W.,  {Gallagher} J.~S.,    {Hunter} D.~A.,  1994, \apj, 433, 65

\bibitem[\protect\citeauthoryear{{Origlia}, {Leitherer}, {Aloisi}, {Greggio} \&
  {Tosi}}{{Origlia} et~al.}{2001}]{origlia01}
{Origlia} L.,  {Leitherer} C.,  {Aloisi} A.,  {Greggio} L.,    {Tosi} M.,
  2001, \aj, 122, 815

\bibitem[\protect\citeauthoryear{{Osterbrock}}{{Osterbrock}}{1989}]{osterbrock%
89}
{Osterbrock} D.~E.,  1989, Astrophysics of Gaseous Nebulae and Active Galactic
  Nuclei.
University Science Books

\bibitem[\protect\citeauthoryear{{Pittard}, {Dyson}, {Falle} \&
  {Hartquist}}{{Pittard} et~al.}{2005}]{pittard05}
{Pittard} J.~M.,  {Dyson} J.~E.,  {Falle} S.~A.~E.~G.,    {Hartquist} T.~W.,
  2005, \mnras, 361, 1077

\bibitem[\protect\citeauthoryear{{Shopbell} \& {Bland-Hawthorn}}{{Shopbell} \&
  {Bland-Hawthorn}}{1998}]{shopbell98}
{Shopbell} P.~L.,  {Bland-Hawthorn} J.,  1998, \apj, 493, 129

\bibitem[\protect\citeauthoryear{{Sidoli}, {Smith} \& {Crowther}}{{Sidoli}
  et~al.}{2006}]{sidoli06}
{Sidoli} F.,  {Smith} L.~J.,    {Crowther} P.~A.,  2006, \mnras, 370, 799

\bibitem[\protect\citeauthoryear{{Slavin}, {Shull} \& {Begelman}}{{Slavin}
  et~al.}{1993}]{slavin93}
{Slavin} J.~D.,  {Shull} J.~M.,    {Begelman} M.~C.,  1993, \apj, 407, 83

\bibitem[\protect\citeauthoryear{{Stil} \& {Israel}}{{Stil} \&
  {Israel}}{1998}]{stil98}
{Stil} J.~M.,  {Israel} F.~P.,  1998, \aap, 337, 64

\bibitem[\protect\citeauthoryear{{Stil} \& {Israel}}{{Stil} \&
  {Israel}}{2002}]{stil02}
{Stil} J.~M.,  {Israel} F.~P.,  2002, \aap, 392, 473

\bibitem[\protect\citeauthoryear{{Suchkov}, {Balsara}, {Heckman} \&
  {Leitherner}}{{Suchkov} et~al.}{1994}]{suchkov94}
{Suchkov} A.~A.,  {Balsara} D.~S.,  {Heckman} T.~M.,    {Leitherner} C.,  1994,
  \apj, 430, 511

\bibitem[\protect\citeauthoryear{{Taylor}, {H{\"u}ttemeister}, {Klein} \&
  {Greve}}{{Taylor} et~al.}{1999}]{taylor99}
{Taylor} C.~L.,  {H{\"u}ttemeister} S.,  {Klein} U.,    {Greve} A.,  1999,
  \aap, 349, 424

\bibitem[\protect\citeauthoryear{{Tokura}, {Onaka}, {Takahashi}, {Miyata},
  {Sako}, {Honda}, {Okada}, {Sakon} et~al.,}{{Tokura} et~al.}{2006}]{tokura06}
{Tokura} D.,  {Onaka} T.,  {Takahashi} H.,  {Miyata} T.,  {Sako} S.,  {Honda}
  M.,  {Okada} Y.,  {Sakon} I.,    et~al., 2006, \apj, 648, 355

\bibitem[\protect\citeauthoryear{{Tomita}, {Ohta} \& {Saito}}{{Tomita}
  et~al.}{1994}]{tomita94}
{Tomita} A.,  {Ohta} K.,    {Saito} M.,  1994, \pasj, 46, 335

\bibitem[\protect\citeauthoryear{{van Dokkum}}{{van
  Dokkum}}{2001}]{vandokkum01}
{van Dokkum} P.~G.,  2001, \pasp, 113, 1420

\bibitem[\protect\citeauthoryear{{Veilleux} \& {Osterbrock}}{{Veilleux} \&
  {Osterbrock}}{1987}]{veilleux87}
{Veilleux} S.,  {Osterbrock} D.~E.,  1987, \apjs, 63, 295

\bibitem[\protect\citeauthoryear{{Waller}}{{Waller}}{1991}]{waller91}
{Waller} W.~H.,  1991, \apj, 370, 144

\bibitem[\protect\citeauthoryear{{Westmoquette}, {Exter}, {Smith} \&
  {Gallagher} III}{{Westmoquette} et~al.}{2007}]{westm07a}
{Westmoquette} M.~S.,  {Exter} K.~M.,  {Smith} L.~J.,    {Gallagher} III J.~S.,
   2007, \mnras, in press, astro-ph/0708.2379, Paper I

\bibitem[\protect\citeauthoryear{{Westmoquette}, {Smith}, {Gallagher} III \&
  {Exter}}{{Westmoquette} et~al.}{2007}]{westm07b}
{Westmoquette} M.~S.,  {Smith} L.~J.,  {Gallagher} III J.~S.,    {Exter} K.~M.,
   2007, \mnras, in press, astro-ph/0708.2682, Paper II

\bibitem[\protect\citeauthoryear{{Westmoquette}, {Smith}, {Gallagher} III,
  {O'Connell}, {Rosario} \& {de Grijs}}{{Westmoquette} et~al.}{2007}]{westm07c}
{Westmoquette} M.~S.,  {Smith} L.~J.,  {Gallagher} III J.~S.,  {O'Connell}
  R.~W.,  {Rosario} D.~J.,    {de Grijs} R.,  2007, \apj, in press, astro-ph/0708.3311

\bibitem[\protect\citeauthoryear{{Zirakashvili} \& {V{\"o}lk}}{{Zirakashvili}
  \& {V{\"o}lk}}{2006}]{zirakashvili06}
{Zirakashvili} V.~N.,  {V{\"o}lk} H.~J.,  2006, \apj, 636, 140

\end{thebibliography}
\bsp

\label{lastpage}
\end{document}